# Corona-Warn-App – Erste Ergebnisse einer Onlineumfrage zur (Nicht-)Nutzung und Gebrauch

Preliminary Results of an Online Survey Evaluating (Non-) Use of the German "Corona-Warn-App"

Jochen Meyer, Thomas Fröhlich, Kai von Holdt

OFFIS – Institut für Informatik, Oldenburg

{meyer|thomas.froehlich|kai.vonholdt}@offis.de

23.11.2020

**Zusammenfassung**

In dieser Studie wird die Corona-Warn-App der Bundesregierung und des Robert-Koch-Instituts mittels einer nicht-repräsentativen Onlineumfrage mit 1482 Teilnehmer*innen auf Gründe der Nutzung und Nicht-Nutzung hin untersucht. Die Studie gibt Einblicke in das Verhalten von Nutzenden der App während der Corona-Pandemie, beleuchtet das Thema Datenschutz und wie die App im Allgemeinen verwendet wird. Unsere Ergebnisse zeigen, dass die App häufig aus Bedenken zum Datenschutz nicht genutzt wird, es aber auch technische Probleme und Sinnzweifel gibt. Zudem wird die App hauptsächlich aus altruistischen Gründen genutzt und häufig geöffnet, um die eigene Risikobewertung anzusehen und das Funktionieren sicherzustellen. Um die Ergebnisse besser zu verstehen, werden diese mit einem Sample von infas 360 mit 10 553 Teilnehmer*innen verglichen. Es zeigt sich, dass sich die Ergebnisse dieser Studie weitgehend auch mit einer größeren Grundgesamtheit vergleichen lassen. Abschießend werden die Ergebnisse diskutiert und Handlungsempfehlungen abgeleitet.

**Abstract**

In this study, the German „Corona-Warn-App" of the German Federal Government and the Robert-Koch-Institute is examined by means of a non-representative online survey with 1482 participants for reasons of use and non-use. The study provides insights into user behavior with the app during the Corona pandemic, highlights the topic of data protection and how the app is used in general. Our results show that the app is often not used due to privacy concerns, but that there are also technical problems and doubts about its usefulness. In addition, the app is mainly used due to altruistic reasons and is often opened to view the own risk assessment and to ensure its functionality. To better understand the results, we compare our results with a sample of infas 360 with 10553 participants. It is shown that the results of this study can be compared to a larger population. Finally, the results are discussed and recommendations for action are derived.

# 1 Einführung

Wir untersuchen, wie und warum Personen die Corona-Warn-App (nachfolgend auch: „CWA" oder – sofern nicht ausdrücklich anders bezeichnet - „die App") nutzen oder nicht, welche Faktoren die Nutzung oder Nicht-Nutzung beeinflussen, und ob die Nutzung der CWA das Verhalten und die Einstellungen zu Corona beeinflusst. Diese Veröffentlichung erfolgt zeitnah nach Abschluss der Onlineumfrage. Sie stellt die vorläufigen Ergebnisse zur Unterstützung des öffentlichen Diskurses zur Verfügung, damit unsere Erkenntnisse und Handlungsempfehlungen als Grundlage für die Verbesserung der CWA genutzt und eine höhere Durchdringung der App in der Bevölkerung erreicht werden können.

Diese Studie wird im Rahmen des Projektes PANDIA , einem vom Bundesministerium für Bildung und Forschung geförderten Forschungsprojekts durchgeführt (FKZ: 16SV8397). Die Studie ist unabhängig von den Herausgebern der Corona-Warn-App, insbesondere dem Robert-Koch-Institut, der SAP AG, und der Telekom.

# 2 Durchführung

## 2.1 Online-Umfrage

Mittels LimeSurvey wurde eine Online-Umfrage erstellt. Dabei gab es keine Einschränkungen im Hinblick auf die Zielgruppe. Der Aufruf zur Teilnahme erfolgte über eine bundesweit gestreute Pressemitteilung sowie über Social-Media-Kanäle des OFFIS. Unseres Wissens nach wurde sie in verschiedenen regionalen und Fachmedien, überwiegend online, aber in keinem großen überregionalen Medium veröffentlicht. Die Teilnahme war vom 17.8. bis 30.9.2020 möglich.

Es wurden folgende Fragen gestellt. Die vollständigen Fragen und Antwortoptionen finden sich im Anhang.

1. Einstiegsfragen
    1.1. Hiermit bestätige ich, dass ich 16 Jahre oder älter bin.
    1.2. Nutzen Sie die Corona-Warn-App?
2. Nur falls 1.2 nicht mit „ja" beantwortet wurde: Gründe der Nichtnutzung
    2.1. Warum nutzen Sie die App nicht oder nicht mehr? (15 Antwortmöglichkeiten, Mehrfachnennung möglich, plus Freitext)
3. Nur falls 1.2 mit „ja" beantwortet wurde: Nutzungsdauer und –Häufigkeit
    3.1. Wie lange nutzen Sie die Corona-Warn-App bereits?
    3.2. Achten Sie darauf, dass Ihre Corona-Warn-App aktualisiert wird?
    3.3. Wie regelmäßig nehmen Sie Ihr Handy mit, wenn Sie das Haus verlassen?
    3.4. Wie oft öffnen Sie die Corona-Warn-App typischerweise auf Ihrem Mobiltelefon?
    3.5. Öffnen Sie die Corona-Warn-App häufiger, wenn sich die äußeren Umstände ändern? (vier Teilfragen mit Abfrage von Zustimmungsgrad plus Freitext)
    3.6. Wenn Sie die Corona-Warn-App auf Ihrem Mobiltelefon öffnen, warum tun Sie das? (fünf Teilfragen mit Abfrage von Zustimmungsgrad plus Freitext)
4. Nur falls 1.2 mit „ja" beantwortet wurde: Nutzungsgründe und -effekte der Corona-Warn-App
    4.1. Wie hat sich Ihre Einstellung zu Ihrem Risiko durch Corona seit Verwendung der Corona-Warn-App geändert? (vier Teilfragen mit Abfrage von Zustimmungsgrad plus Freitext)
    4.2. Welche Beweggründe haben Sie für die Verwendung der Corona-Warn-App? (sechs Teilfragen mit Abfrage von Zustimmungsgrad plus Freitext)
        4.2.1. Sonstiges: (Freitext)
    4.3. Was ist Ihnen zum Thema Datenschutz der Corona-Warn-App bekannt? (fünf Teilfragen mit Abfrage von Zustimmungsgrad)

5. Demografische Angaben (für alle Teilnehmer)
    5.1. Wie alt sind Sie?
    5.2. Welchem Geschlecht ordnen Sie sich zu?
    5.3. In welchem Land haben Sie Ihren Hauptwohnsitz?
    5.4. Nur falls bei 5.3 „Deutschland" ausgewählt wurde: Bitte geben Sie die ersten 3 Stellen der Postleitzahl Ihres Wohnortes an.
    5.5. Was ist ihre höchste, abgeschlossene Ausbildung?
    5.6. Benutzen Sie andere Tracking-Apps und Geräte?
    5.7. Wie ist - unabhängig von der Corona Warn-App – Ihre persönliche Einstellung
        5.7.1. zu Datenschutz im Allgemeinen (2 Fragen),
        5.7.2. zur Bedrohung für Sie durch Corona und (2 Fragen)
        5.7.3. zu neuen Technologien im Allgemeinen (2 Fragen)?

Bei den Abfragen des Zustimmungsgrads (in 3.5, 3.6, 4.1, 4.2, 4.3, 5.7) wurden die Antworten auf einer 5-Punkt-Skala „Stimme gar nicht zu" – „Stimme eher nicht zu" – „Neutral" – „Stimme eher zu" – „Stimme voll und ganz zu" erfasst. Nachfolgend werden diese auf die Skala -2 bis +2 abgebildet, sodass Antworten im negativen Bereich ablehnend, bei 0 neutral, und im positiven Bereich zustimmend sind. Bei statistisch signifikanten Ergebnissen werden die folgenden Codes für die Kennzeichnung des p-Werts verwendet: 0 '***', 0.001 '**', 0.01 '*'. In den folgenden statistischen Auswertungen wurden Bonferroni-Korrekturen durchgeführt, die sich auf die Anzahl der jeweiligen p-Werte beziehen und nicht auf die Gesamtzahl aus allen Tests.

Die Fragen zu 5.7 basieren in Anlehnung an (Trang, et al. 2020) auf (Agarwal, Malhotra und Kim 2004) für Datenschutzbedeutung, auf (Heinssen, Glass und Knight 1987) für Technikaffinität und auf (Salkovskis, et al. 2002) zur Corona-Sorge. Als Indikator wurde pro Teilnehmer der Mittelwert der jeweiligen Antworten ermittelt, um die drei Themen Datenschutz, Corona-Sorge und Technikaffinität als persönliche Faktoren zu identifizieren.

## 2.2   infas 360-Survey

Parallel zu dieser Studie wurde von der infas 360 GmbH eine größere, nicht repräsentative Online-Umfrage zu Corona und der Warn-App durchgeführt. In Abstimmung mit infas 360 wurden die folgenden Fragen unserer Studie unverändert in die infas 360-Umfrage übernommen:

- 1.2 Nutzen Sie die Corona-Warn-App?
- 2.1 Warum nutzen Sie die App nicht oder nicht mehr?
- 3.3 Wie regelmäßig nehmen Sie Ihr Handy mit, wenn Sie das Haus verlassen?
- 3.4 Wie oft öffnen Sie die Corona-Warn-App typischerweise auf Ihrem Mobiltelefon?
- 3.6 Wenn Sie die Corona-Warn-App auf Ihrem Mobiltelefon öffnen, warum tun Sie das?
- 4.2 Welche Beweggründe haben Sie für die Verwendung der Corona-Warn-App?

Die Einbindung dieses Teils unserer Fragen in einen größeren Kontext ermöglicht es abzuschätzen, inwiefern sich unser Sample und damit die Ergebnisse dieser Studie mit einer größeren Grundgesamtheit vergleichen lassen. Zusätzlich zu diesen Fragen wurden in der infas 360-Studie noch weitere Fragen in Bezug auf die Corona-Warn-App gestellt. Diese sind aber in der vorliegenden Arbeit nicht berücksichtigt.

# 3   Beschreibung der Samples

## 3.1   OFFIS-Sample

Es haben 1482 Personen an der Umfrage teilgenommen. Von denen haben 154 die Umfrage nicht abgeschlossen. Dabei sind *n = 1326* Antworten sind vollständig und werden nachfolgend als Grundgesamtheit be-rücksichtigt. Dabei ist zu bemerken, dass sich bei der statistischen Auswertung einzelner Fragen und Zusammenhänge veränderliche Teilmengen aus der Grundgesamtheit ergeben haben. Das begründet sich mit dem Ausschluss von Antworten die die Angaben „keine Angabe" und „weiß nicht" enthalten. So wurden bei der Auswertung aus der Grundgesamtheit die jeweils maximal große Teilmenge von gültigen Antworten für jede Analyse ermittelt, die die beiden ausschließenden Antwortoptionen gleichermaßen nicht enthielten und somit nur inhaltlich vollständige Antworten abbilden. Das veränderliche *n* der Teilmengen wird im Folgenden mit angegeben.

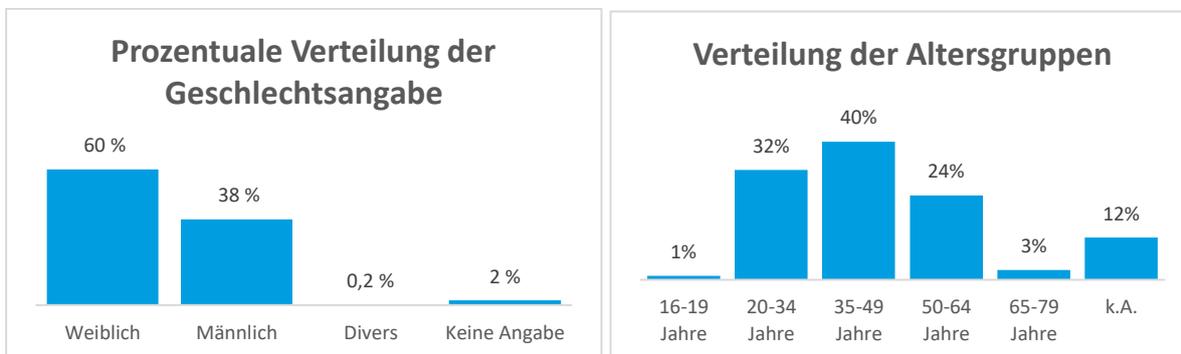

Frauen waren mit 60 % (*n=795*) stärker vertreten (502 männlich, 2 divers, 27 keine Angabe). Das Durchschnittsalter der Teilnehmenden lag mit 42,2 Jahren etwas unter dem Bundesdurchschnitt von 44,5 Jahren[1], wobei jüngere und ältere Personen unterrepräsentiert sind. Der Bildungsstand der Teilnehmende war erheblich höher als im Bundesdurchschnitt (ca. 30 % haben im Bundesdurchschnitt ein Studium abgeschlossen, eine Promotion haben ca. 1,4 % abgeschlossen[2]). Insbesondere haben zusammen 65 % der Teilnehmenden (*n=860*) ein Studium oder eine Promotion abgeschlossen (10 Hauptschule, 129 mittlere Reife, 267 Abitur, 20 sonstiges, 40 keine Angabe).

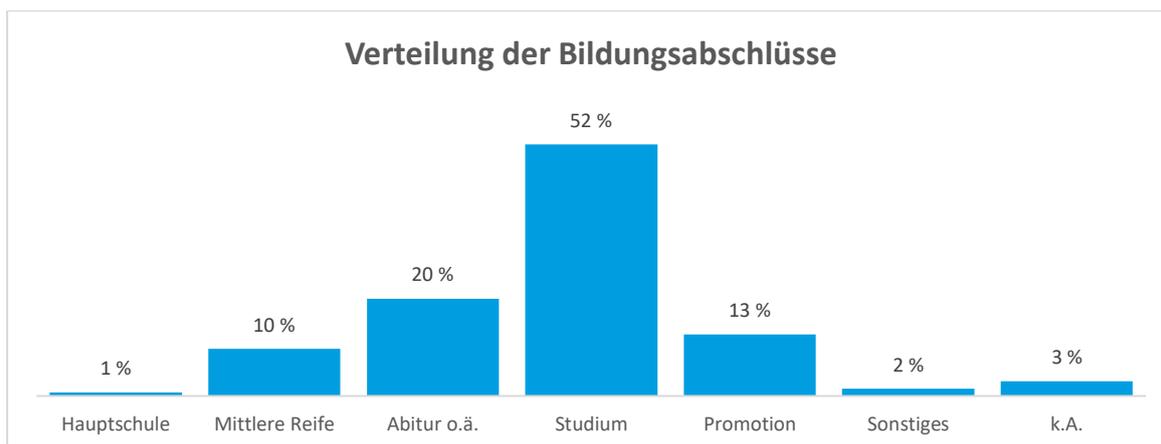

Es haben wesentlich mehr Nutzende der App (*n=1079*, 81 %) als Nicht-Nutzende (*n=246*, 19 %) teilgenommen. Die Gruppen unterscheiden sich nicht wesentlich hinsichtlich Geschlechtes, Altersverteilung oder Bildungsabschluss. Die Verteilung der Antworten über Deutschland ist ungleichmäßig. Das Gebiet 2 (n=278), und speziell 261 (Oldenburg (n=90)) ist überdurchschnittlich vertreten. Es sind aber aus jeden Postleitzahlengebiet 0-9 mindestens jeweils 50 Antworten eingegangen. 6 Antworten kamen aus dem Ausland (Belgien 3; Niederlande, Schweiz, Polen je 1).

---

[1] Quelle: https://de.statista.com/statistik/daten/studie/723069/umfrage/durchschnittsalter-der-bevoelkerung-in-deutschland-nach-staatsangehoerigkeit/
[2] Quelle: https://www.sueddeutsche.de/bildung/oecd-bildung-bundeslaender-akademiker-1.4596188

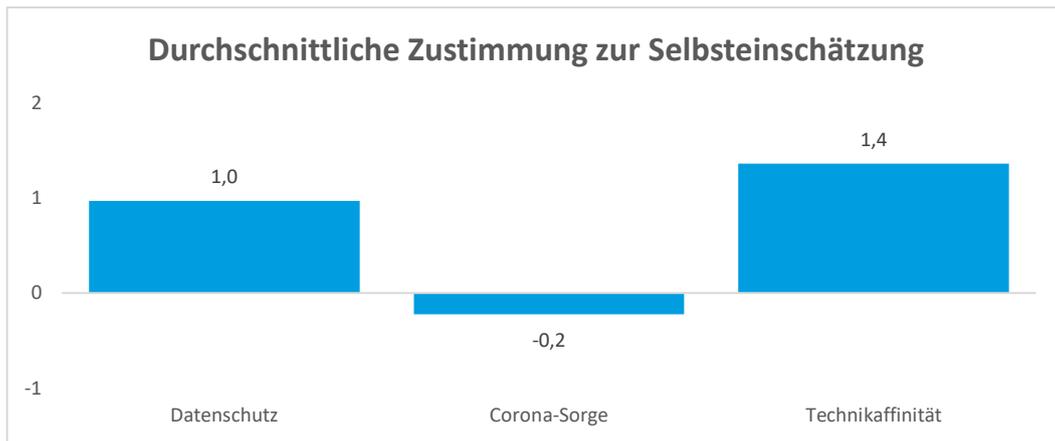

Wir fragten nach der Selbsteinschätzung in den Bereichen Datenschutz, Corona-Sorge und Technikaffinität (Fragen 5.7). Die Bedeutung von Datenschutz war mit 1,0 (auf einer Skala von -2 bis 2) eher moderat, die Sorge vor Corona mit -0,2 nur leicht ausgeprägt. Technologieaffinität war mit 1,4 deutlich hoch. Diese Werte wurden durch eine anteilige Gewichtung der Durchschnittwerte der ursprünglichen sechs Selbsteinschätzungsfragen ermittelt.

### 3.2 infas 360 Sample

Das von infas 360 erhobene Sample, welches wir zum Vergleich heranziehen, ist mit 10553 Teilnehmern erheblich größer und hinsichtlich Geschlechts- (50,6 % männlich) und Altersverteilung näher am Bundesschnitt, aber ebenfalls nicht repräsentativ. Der Anteil der Nicht-Nutzenden ist mit 62 % (*n=6.549*) erheblich größer als im OFFIS-Sample. Eine ausführlichere Beschreibung erfolgt in Abschnitt 5.

## 4 Ergebnisse der OFFIS-Umfrage

Einleitend weisen wir darauf hin, dass die im Folgenden präsentierten statistischen Analysen einer explorativen Vorgehensweise zugrunde liegen. Das bedeutet, dass statistische Auswertungen, insbesondere Korrelationen und statistische Unterschiede auf Basis von informierten Annahmen durchgeführt wurden und keine Hypothesenbildung vor der Auswertung durchgeführt wurde. Dementsprechend sind die Ergebnisse indikativ zu bewerten und beschreiben lediglich das gemeinsame Auftreten zweier Faktoren, nicht jedoch kausale Zusammenhänge. Fehlerindikatoren in den folgenden Grafiken bilden das Konfidenzintervall mit einem Alpha-Wert von 0,05 ab.

### 4.1 Nicht-Nutzung und Vergleich zu Nutzenden

Von allen Teilnehmenden haben 11 % die App nie installiert, und jeweils 2-3% haben erfolglos versucht sie zu installieren, hatten sie installiert aber nicht oder nicht mehr aktiviert, oder haben sie wieder deinstalliert.

Dominierende Gründe für Nicht-Nutzung sind Datenschutzbedenken, Zweifel, dass die App sinnvoll ist, und Probleme mit der Technik (z.B. veraltetes Handy, zu wenig Speicher, Probleme mit dem Akku), außerdem Probleme mit der User Experience (z.B. irritierende Fehlermeldungen, Langeweile der App, nicht erkennen was die App macht). Noch sichtbar, aber nachgeordnet waren Gleichgültigkeit und Zweifel an der Bedrohung durch Corona. Kaum Probleme bestehen beim Zugang zur App.

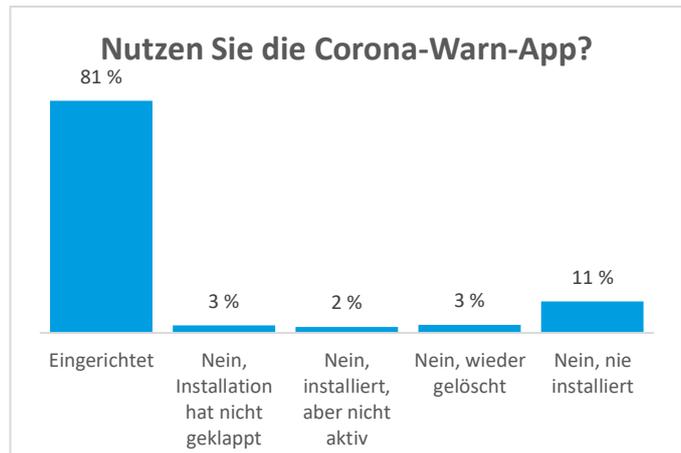

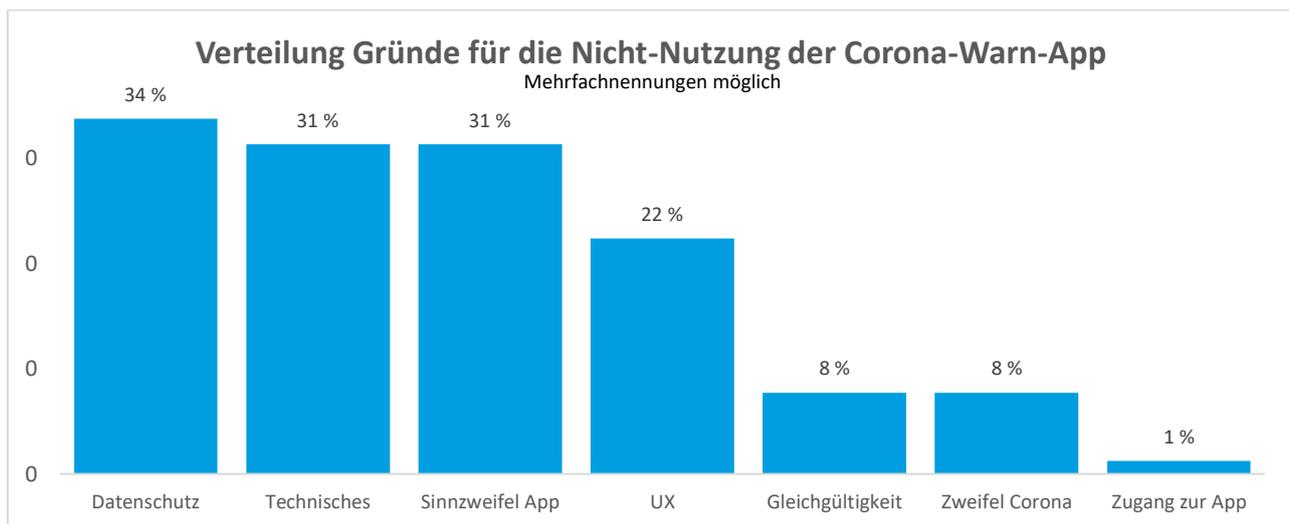

Im Vergleich zu den Nutzenden war den Nicht-Nutzenden der Datenschutz leicht, aber statistisch signifikant wichtiger (Nutzende: 0,72; Nicht-Nutzende: 0,93; p = 0,002371). Die Sorge um Corona war in beiden Gruppen eher gering, aber bei Nicht-Nutzenden statistisch signifikant geringer (Nutzende: -0,39; Nicht-Nutzende: -0,72, p < 0,0005). Die Fehlerindikatoren in allen folgenden Grafiken bilden das Konfidenzintervall ab.

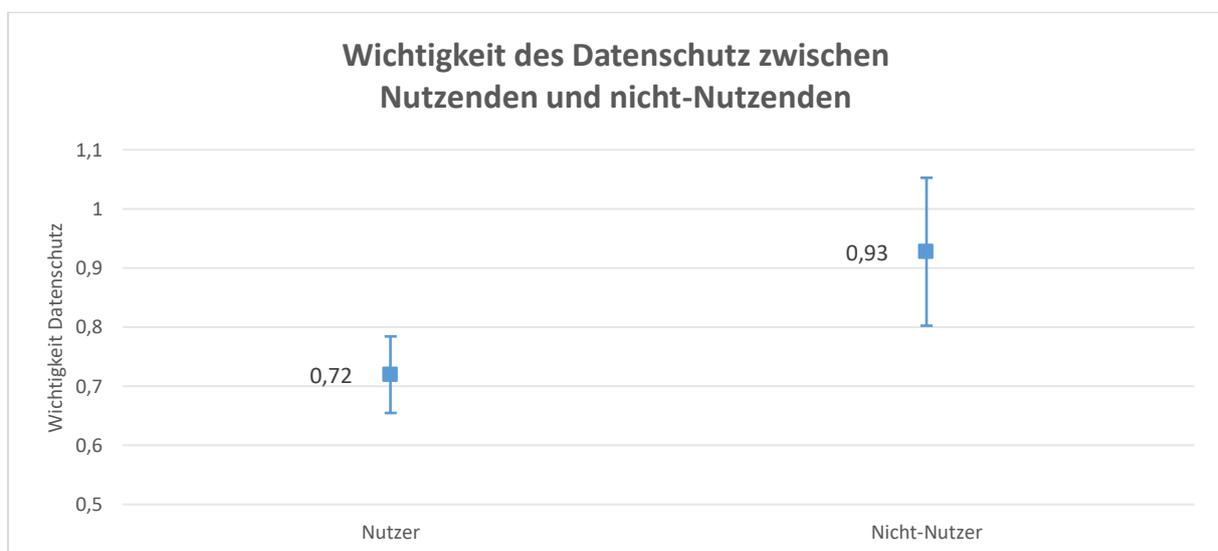

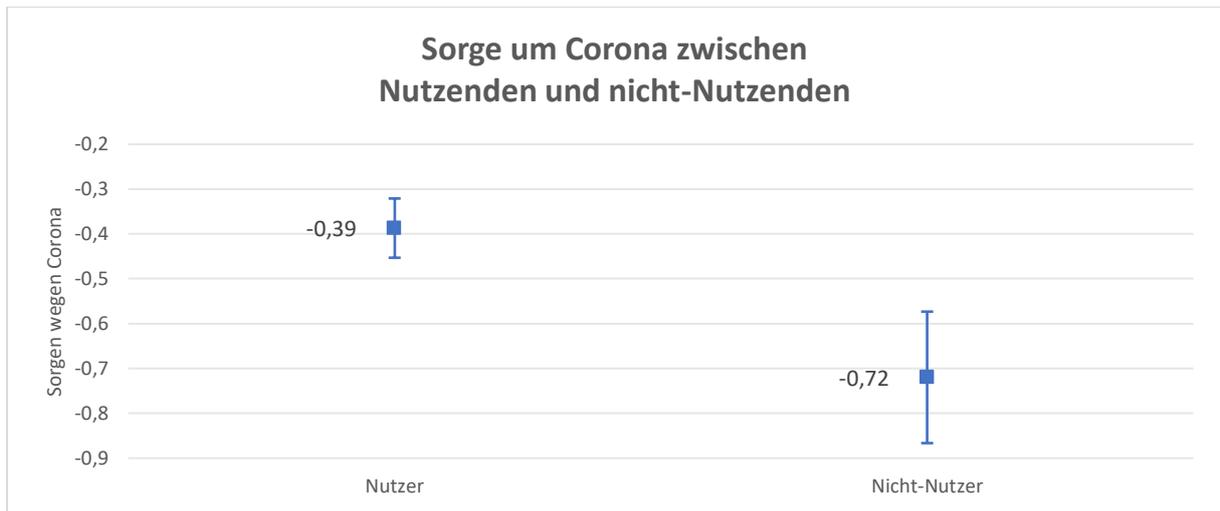

Nutzende wie auch Nicht-Nutzende verwenden auch andere vernetze Tracking-Geräte. Dabei spielen Apps und Aktivitätstracker mit insgesamt 40 % aktiver Nutzung (Nicht-Nutzende 29 %) eine große Rolle. Sportuhren und Apps werden insgesamt von 34 % aktiv genutzt (Nicht-Nutzende 22 %). Smartwatches werden von insgesamt 21 % aktiv verwendet (Nicht-Nutzende 15 %).

## 4.2 Art der Nutzung der CWA

### 4.2.1 Nutzungsdauer und Maintenance

Der überwiegende Anteil der Nutzenden (67 %, *n=892*) hat die App seit ihrer Veröffentlichung installiert. Weniger als 4 % (*n=47*) nutzen sie seit bis zu einem Monat. Die Grundvoraussetzung für die Nützlichkeit der App, das Handy regelmäßig mit sich zu führen, erfüllen 71 % (*n=939*) immer, und 9 % (*n=120*) oft. Nur eine Minderheit von 1 % hat das Handy gelegentlich oder selten dabei; niemand gab an, das Handy nie dabei zu haben.

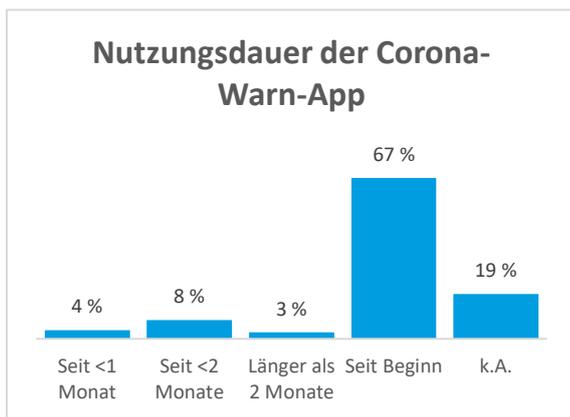

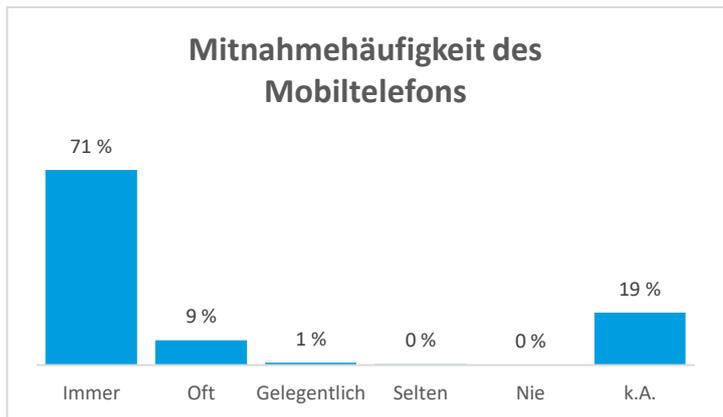

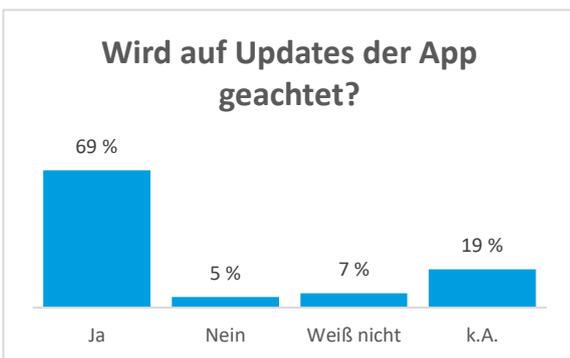

Die Aktualität der App bei den Nutzenden ist insgesamt sehr gut: Der weit überwiegende Teil (69 %, *n=910*) der Teilnehmer achtet darauf, dass die App regelmäßig aktualisiert wird. Aber immerhin 5 % verneinten dies, und 7 % gaben an, nicht zu wissen, ob sie sie regelmäßig aktualisieren. Hier kann vermutet werden, dass bei zumindest einem Anteil der Nutzenden dies regelmäßig automatisch im Hintergrund als Teil der Systemeinstellungen des jeweiligen Handys passiert.

### 4.2.2 Umgang mit der CWA
#### *4.2.2.1 Häufigkeit und Gründe des Öffnens der App*

Wir fragten, wie oft und wieso Nutzende die CWA öffnen, d.h. sie vom Startbildschirm des Handys aus aufrufen. Es ist zu beachten, dass das Öffnen der App nur aus wenigen Gründen tatsächlich notwendig ist, beispielsweise zum Eingeben eines Testergebnisses oder zur Sicherstellung des Funktionierens der Corona-Warnung, was aufgrund eines mittlerweile behobenen Bugs zwischenzeitlich notwendig erschien. Zum Zeitpunkt der Umfrage wurde nicht aktiv dazu aufgefordert, die App regelmäßig zu öffnen, um den Corona-Status zu überprüfen. Zum aktuellen Zeitpunkt (Mitte November 2020) wird aktuell von der CWA in regelmäßiger Weise eine Benachrichtigung angezeigt, die dazu auffordert, den Corona-Status in der App zu prüfen. Die weiteren Funktionen der App, auch das Warnen zum Corona-Status laufen im Prinzip automatisch bzw. im Hintergrund. Die überwiegende Mehrheit von 67 % (*n=886*) der Nutzenden öffnet die CWA dennoch regelmäßig mehrmals wöchentlich oder täglich. Nur unter 15 % (*n=192*) öffnen sie seltener als einmal im Monat oder nie.

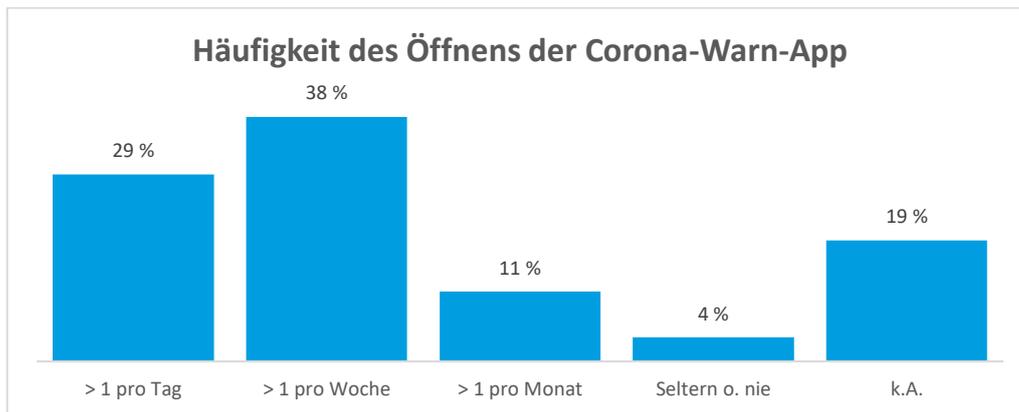

Im Zusammenhang mit der aktiven Nutzung der App stellte sich die Frage inwiefern die persönlichen Faktoren Datenschutz, Sorge um Corona und Technikaffinität hiermit zusammenhängen, da angenommen werden kann, dass Personen mit höherer Sorge um Corona auch häufiger die App aktiv nutzen, um beispielsweise ihren Corona-Status zu überprüfen. Wir führten deshalb eine Rangkorrelationsanalyse nach Spearman durch und fanden, dass die Sorge um Corona einen negativ statistisch signifikanten Zusammenhang zeigt mit *ρ = -0,18 und p = 5,9e-07*:

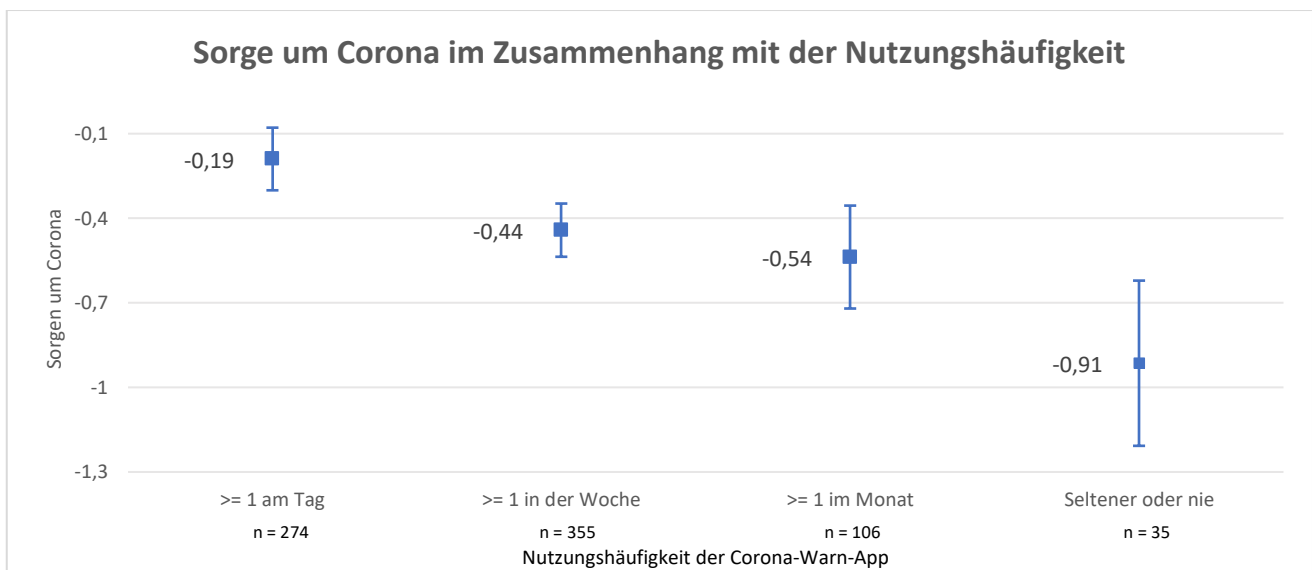

Um festzustellen, welche Gruppen sich hier unterscheiden, führten wir einen paarweisen Vergleich nach Conover mit Bonferroni-Korrektur durch und fanden signifikante Unterschiede zwischen der Gruppe 1 (>= 1 am Tag") und allen restlichen Gruppen mit *p < 0,005* und Gruppe 2 (>= 1 in der Woche) und 4 (Seltener oder nie) mit *p < 0,05*. Die gleiche Analyse zeigte zudem eine Korrelation mit *ρ = 0,13 und p < 0,0005* für den Faktor Technikaffinität. Ein paarweiser Vergleich ergab signifikante Unterschiede zwischen den Gruppe 1 und den Gruppen 2 und 3 mit *p < 0,05*:

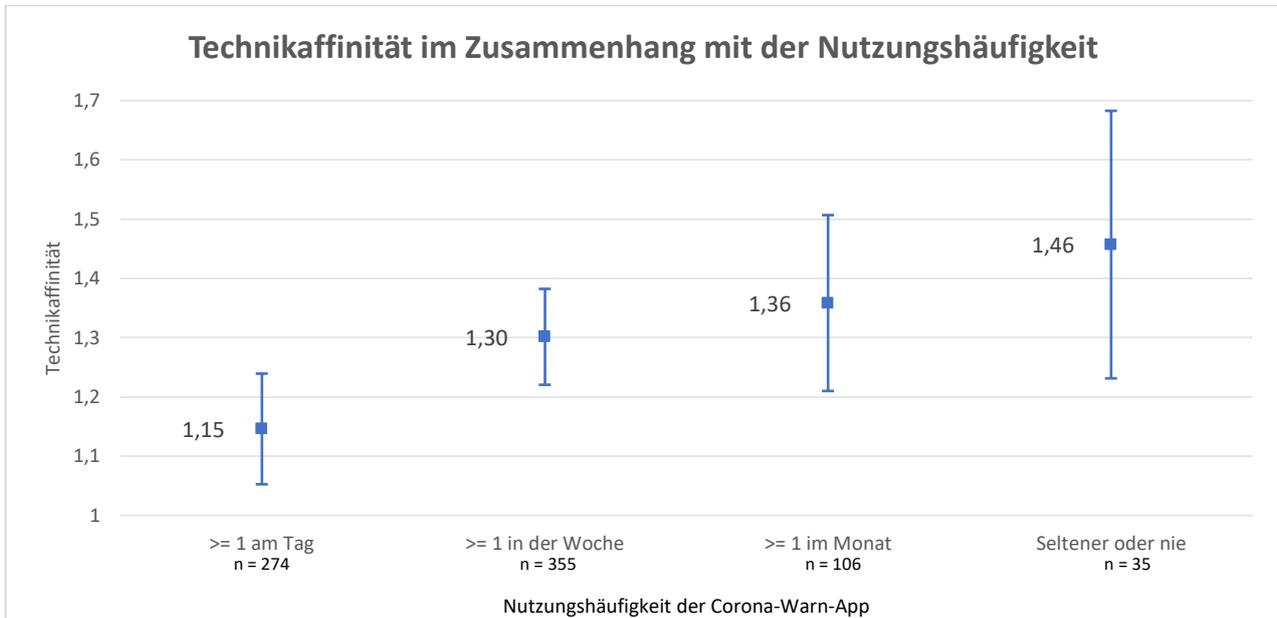

Die dominierenden Gründe für das Öffnen der App sind mit starker Zustimmung das Überprüfen des eigenen Corona-Status (Durchschnitt 1,4), und die Überprüfung des technischen Funktionierens der App (1,2). Hinsichtlich des Eingebens eines positiven oder negativen Testergebnisses gab es zwar ebenfalls eine deutliche Zustimmung (1), aber einen erheblichen Anteil von „weiß-nicht"-Angaben (20,2 %). Neutral war der Grund, die App aus allgemeinem Interesse zu öffnen (-0,03), und deutlich ablehnend zum Erhalten allgemeiner Corona-Informationen (-1,1).

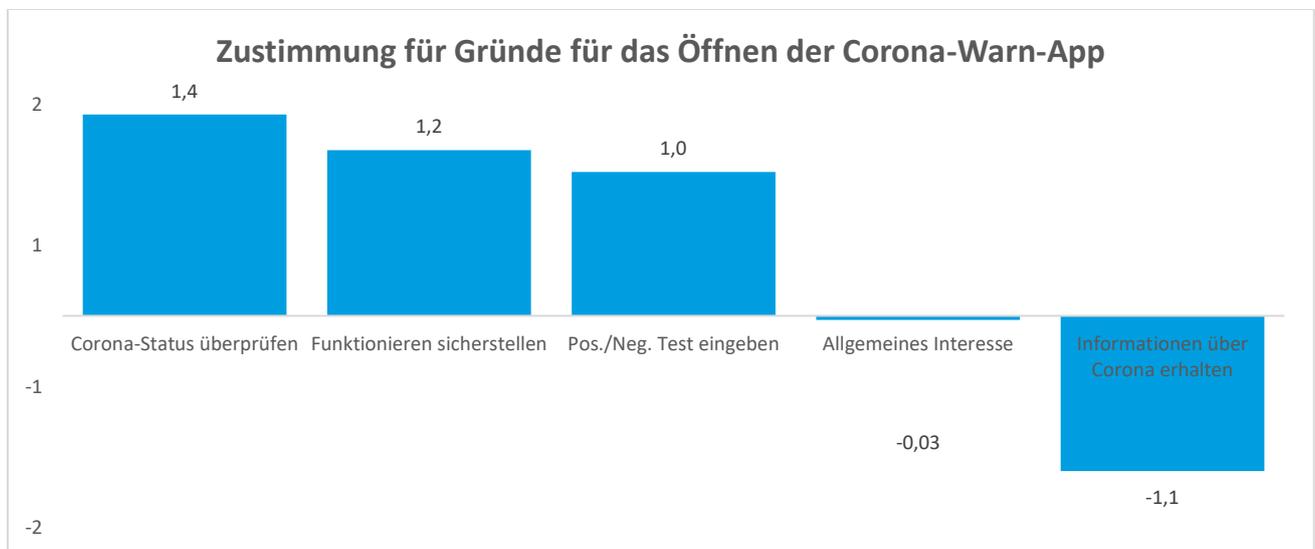

Eine Korrelation gab es insbesondere zur Corona-Sorge: Personen mit höherer Corona-Sorge öffnen die App etwas häufiger, um ihren Corona-Status zu überprüfen und um das Funktionieren der App sicherzustellen, und sind auch weniger ablehnend darin, Informationen über Corona aus der App zu erhalten. Letzteres gilt auch für Personen mit höherem Datenschutzinteresse, wobei hier der logische Zusammenhang nur schwer herzustellen ist, bzw. zufällig sein kann.

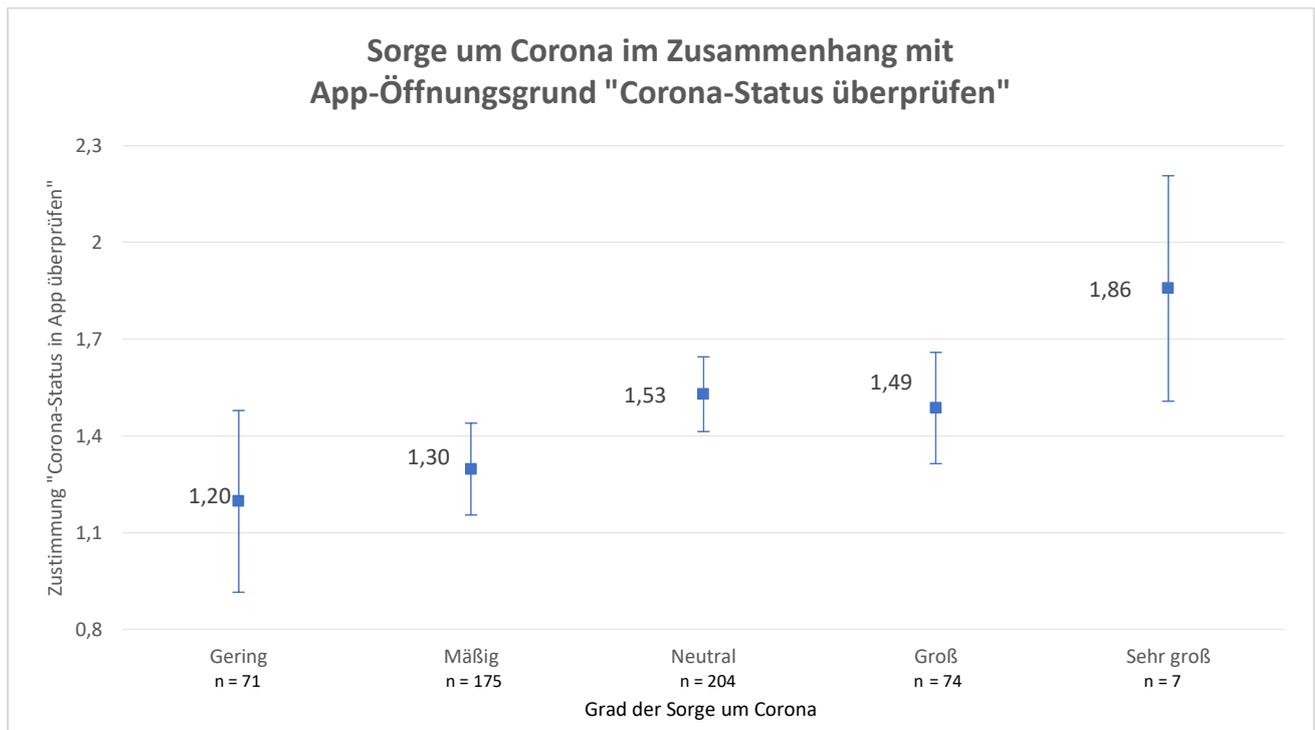

Spearman-Korrelation mit *ρ = 0,12 und p < 0,05*. Signifikante Unterschiede zwischen den Gruppen „Mäßig" und „Neutral" nach Conover-Test mit Bonferroni-Korrektur mit *p < 0,05*.

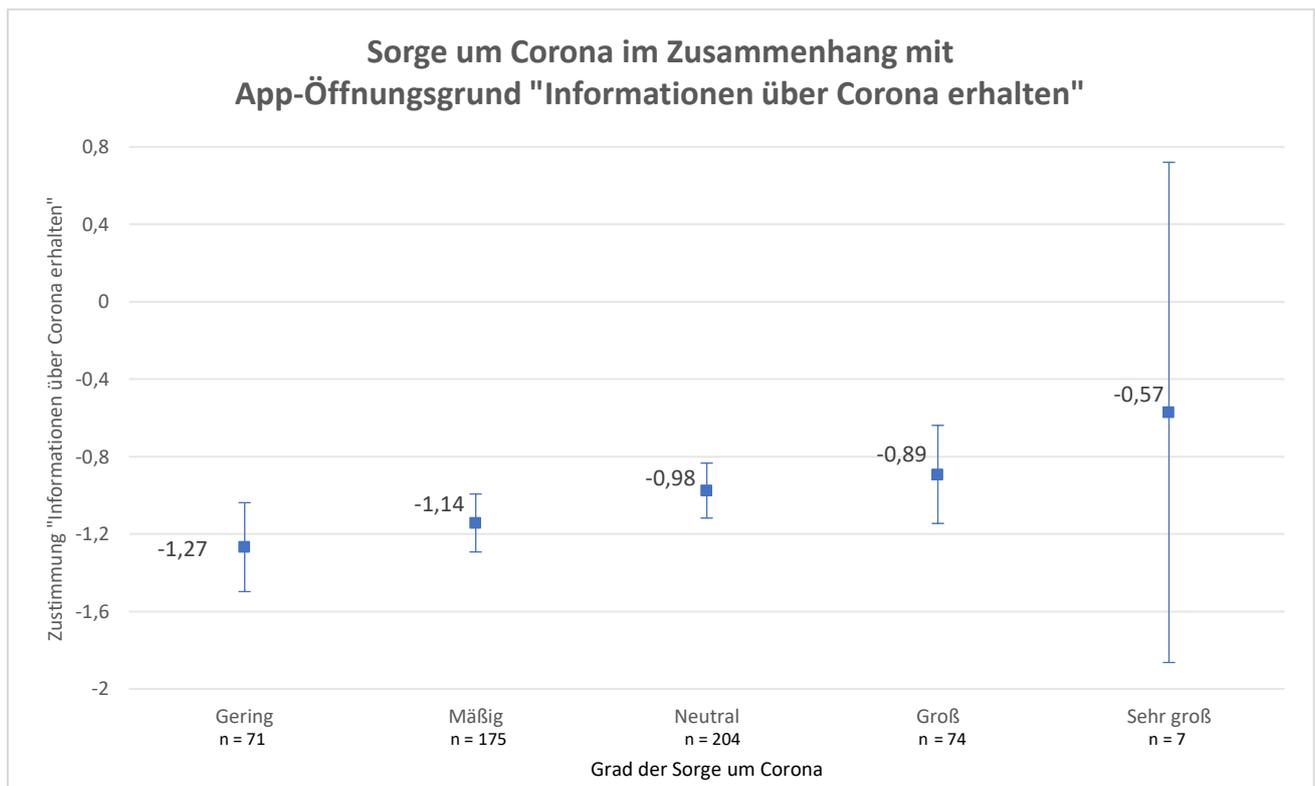

Spearman-Korrelation mit *ρ = 0,13 und p < 0,005*. Keine signifikanten Unterschiede zwischen den Gruppen.

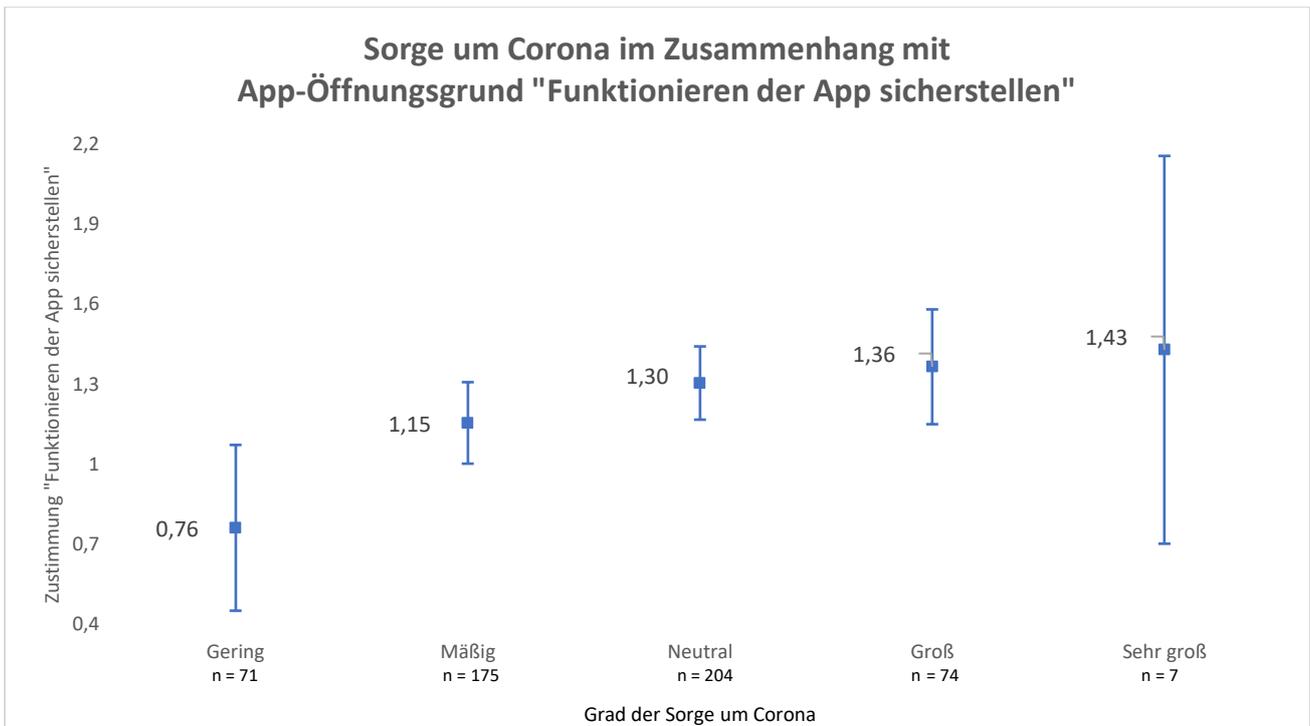

Spearman-Korrelation mit $\rho = 0{,}16$ und $p < 0{,}0005$. Signifikante Unterschiede zwischen den Gruppen „Gering" und „Neutral" nach Conover-Test mit Bonferroni-Korrektur mit $p < 0{,}005$. Zudem signifikante Unterschiede zwischen den Gruppen „Gering" und „Groß" mit $p < 0{,}05$.

Bei Betrachtung der Altersgruppen in Bezug auf die Aussage, dass die App genutzt wird, um Informationen über Corona zu erhalten wird deutlich, dass in der Altersgruppe der „Ab-59"-jährigen die Ablehnung, Informationen über Corona aus der App zu erhalten, weniger stark ausgeprägt ist.

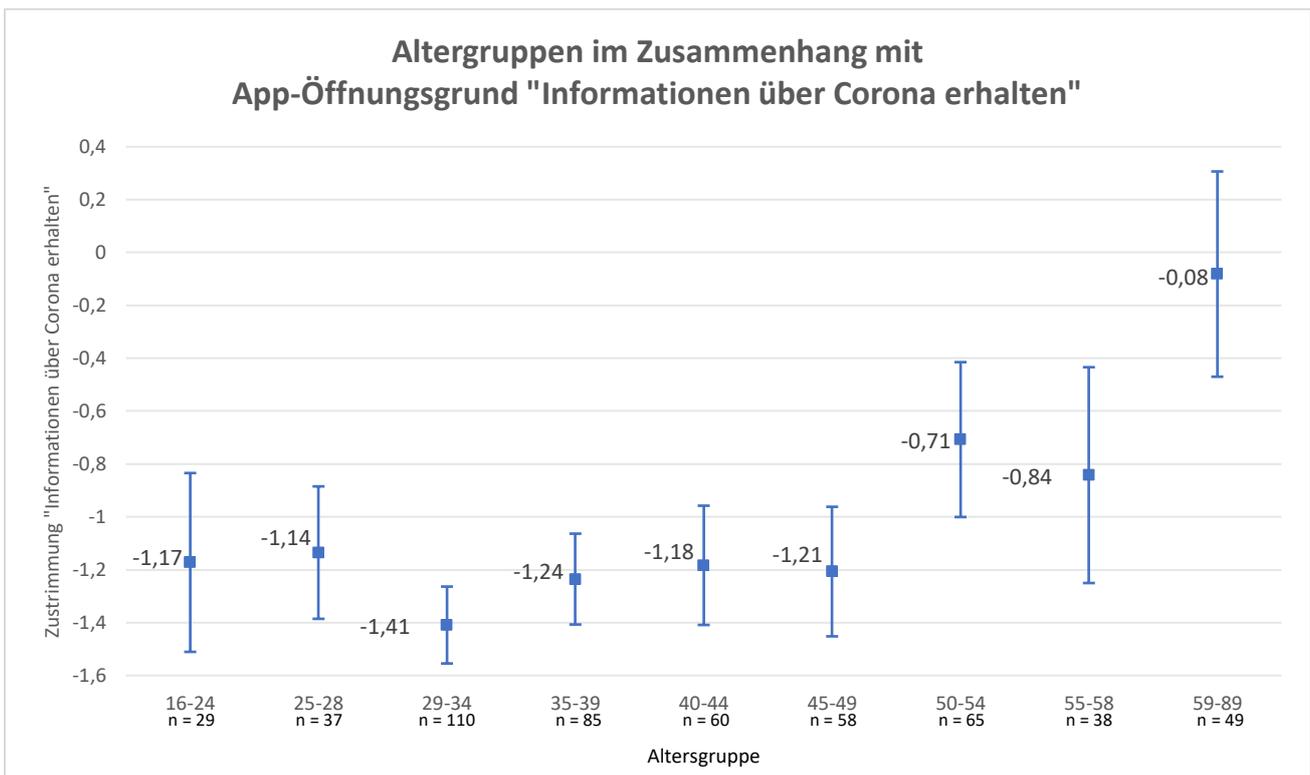

Spearman-Korrelation mit $\rho = 0{,}23$ und $p = 8{,}164e-08$. Signifikante Unterschiede zwischen den Gruppen (in der Grafik von links nach rechts) {1*,2**,3***,4***,5***,6***,7***} und {9} nach paarweisem Wilcoxon-Rangsummentest mit Bonferroni-Korrektur.

*4.2.2.2 Vermutete Veränderungen der Öffnungshäufigkeit [App-Öffnungs-Umstände]*

Auf die Frage, ob sie die CWA bei Veränderungen der äußeren Umstände häufiger öffnen würden, gab es eine Zustimmung in den Situationen, dass es im sozialen Umfeld einen Fall gibt (1,13) und dass es in der Nähe einen Ausbruchsherd gibt (Durchschnitt 1,06). Im letzteren Fall war der Anteil der Unsicheren, die mit „weiß nicht" geantwortet haben, mit knapp 10 % etwas höher als bei den anderen Fragen. Kaum Zustimmung (0,42) gab es für den Fall, dass sich die Corona-Situation in Deutschland insgesamt verschlechtert, und eine leichte Ablehnung (-0,25), wenn Corona in den Medien präsenter ist.

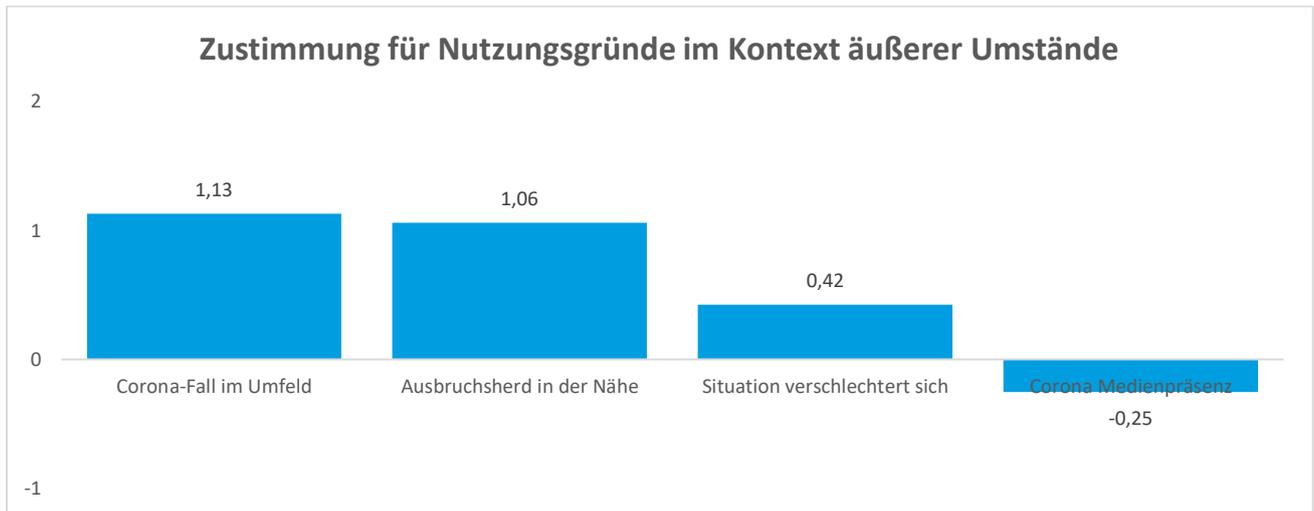

Hier zeigt sich ein Zusammenhang zu der Corona-Sorge: Eine höhere Corona-Sorge ist korreliert mit einem häufigeren Öffnen, wenn sich die Corona-Situation in Deutschland insgesamt verschlimmert – hier mit einem starken Effekt: Sind Personen mit geringer Corona-Sorge hier neutral (-0,02), stimmen sehr besorgte Personen deutlich zu (1,67).

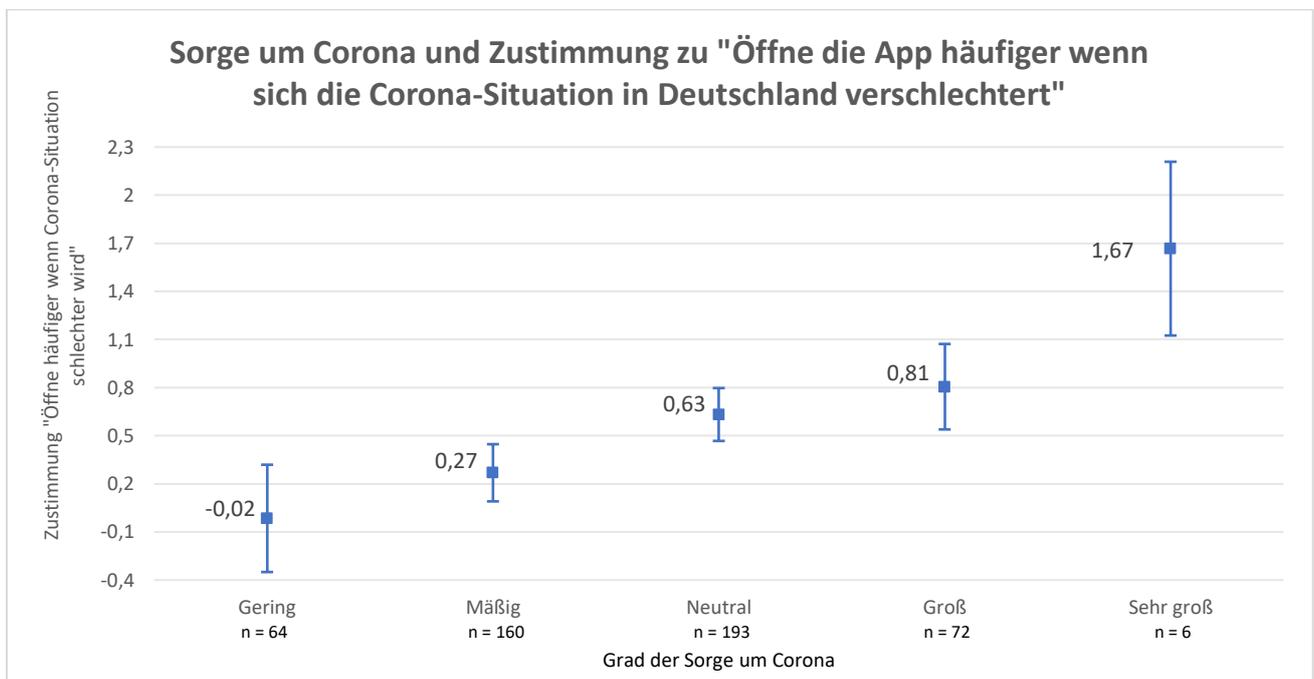

Spearman-Korrelation *mit ρ = 0,24 und p = 7,8e-08*. Signifikante Unterschiede ergeben sich zwischen den Gruppen {3**,4**,5**} und {1} und {3*,4**,5*} und {2} nach einem paarweisen Vergleich nach Conover mit Bonferroni-Korrektur. Wenn es einen Ausbruchsherd in der Nähe gibt, ist der Effekt moderater. Von moderater (0,86) auf sehr deutliche Zustimmung (1,5), ebenso, wenn Corona in den Medien präsenter ist (von leichter Ablehnung (-0,53) auf moderate Zustimmung (0,83)). Keinen Zusammenhang gibt es, wenn es Fälle im sozialen Umfeld gibt. In dem Zusammenhang ist aber die Zustimmung unabhängig von der Corona-Sorge moderat hoch (1,13, siehe oben).

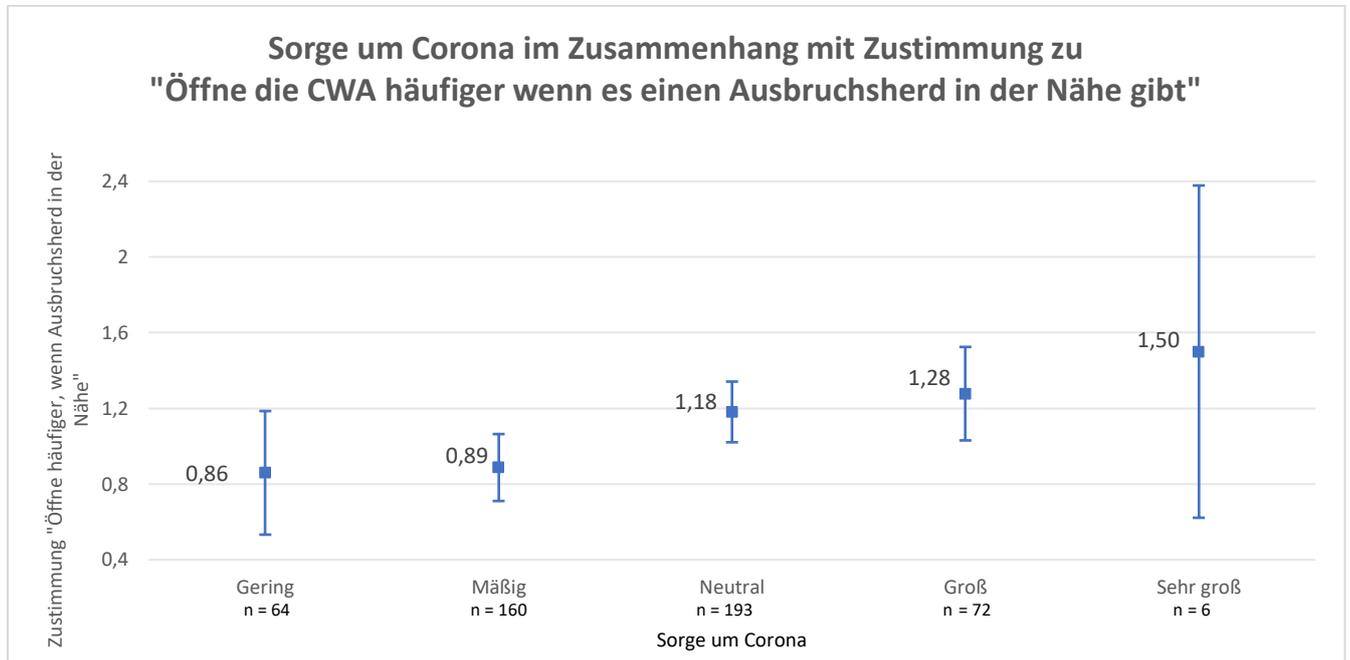

Korrelation nach Spearman mit *ρ = 0,16 und p < 0,0005*. Signifikante Unterschiede ergeben sich zwischen den Gruppen {3*,4*} und {2} nach paarweisem Vergleich nach Conover mit Bonferroni-Korrektur.

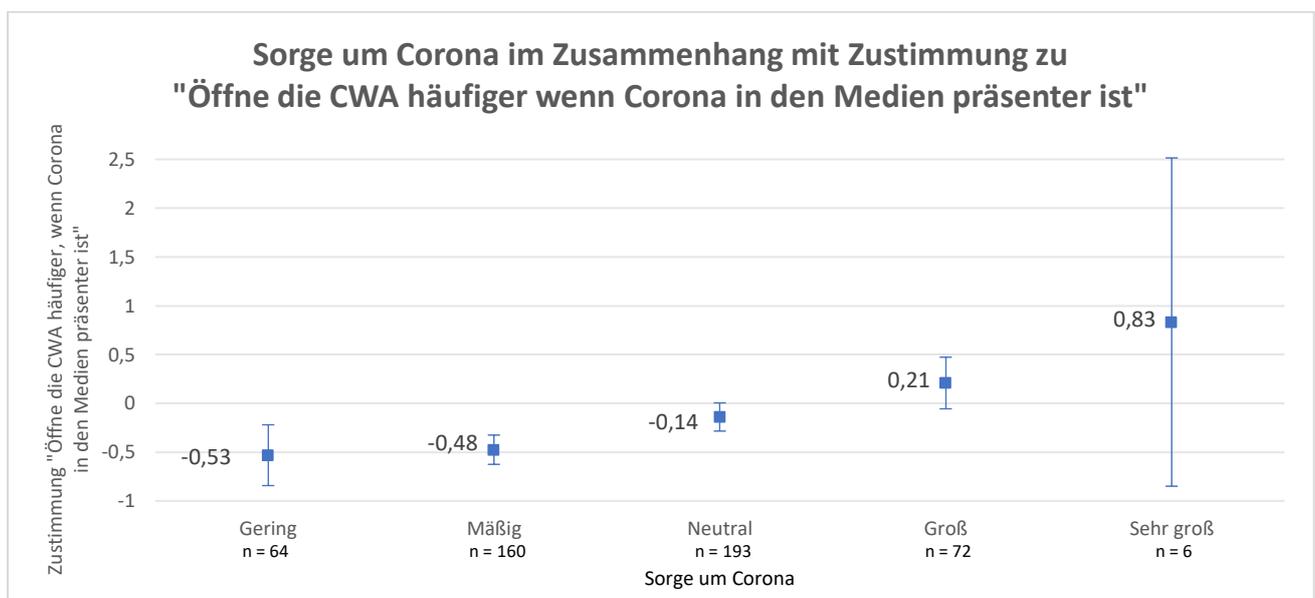

Korrelation nach Spearman mit *ρ = 0,24 und p = 4,5e-08*. Signifikante Unterschiede zwischen den Gruppen 4*** und 1 und den Gruppen {3*,4***} und 2 nach paarweisem Vergleich nach Conover mit Bonferroni-Korrektur.

## 4.3 Gründe für die Nutzung der CWA

Der dominierende Grund für die Nutzung der App ist der altruistische Wunsch, zum Beenden der Pandemie beizutragen (1,6). Ebenfalls Zustimmung fand der Schutz der eigenen Familie (1,0). Nur noch geringe Zustimmung gab es zu dem Grund, sich selber zu schützen (0,3). Eher abgelehnt wurden die extrinsisch motivierten Gründe, von Freunden oder Medien überzeugt oder gedrängt worden zu sein.

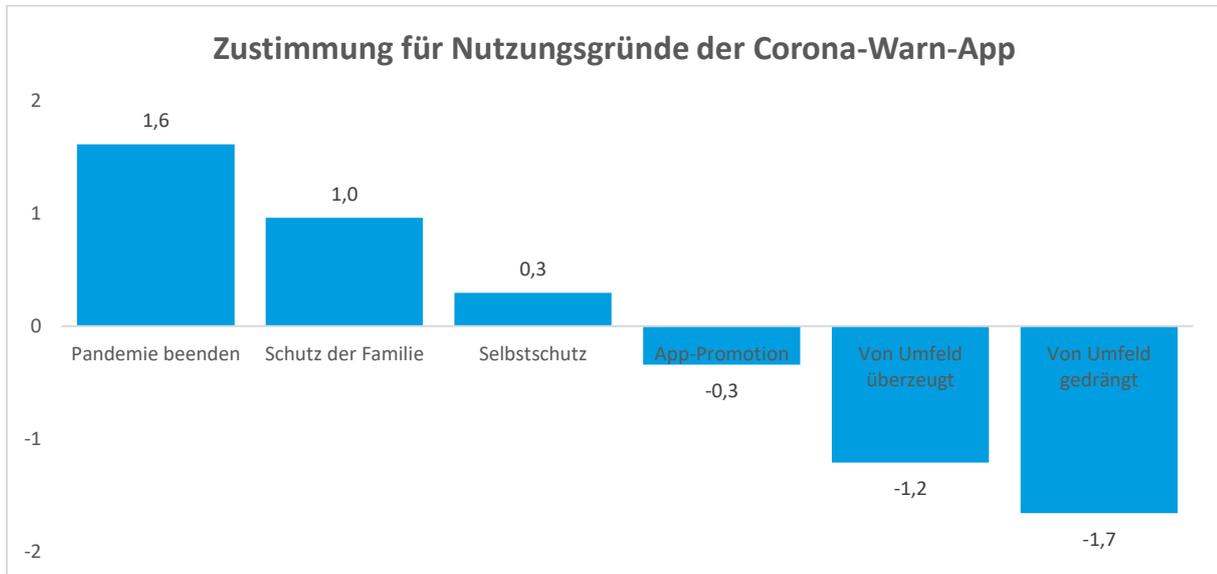

Bei Personen mit höherer Corona-Sorge steigt die Zustimmung zum Familien- und zum Selbstschutz sowie zum Pandemie-Beenden stärker an. Auch Medien und Werbung tragen hier stärker zur Nutzung bei, nicht jedoch die Überzeugung durch Freunde und Verwandte. Sie fühlten sich aber etwas stärker durch Dritte gedrängt.

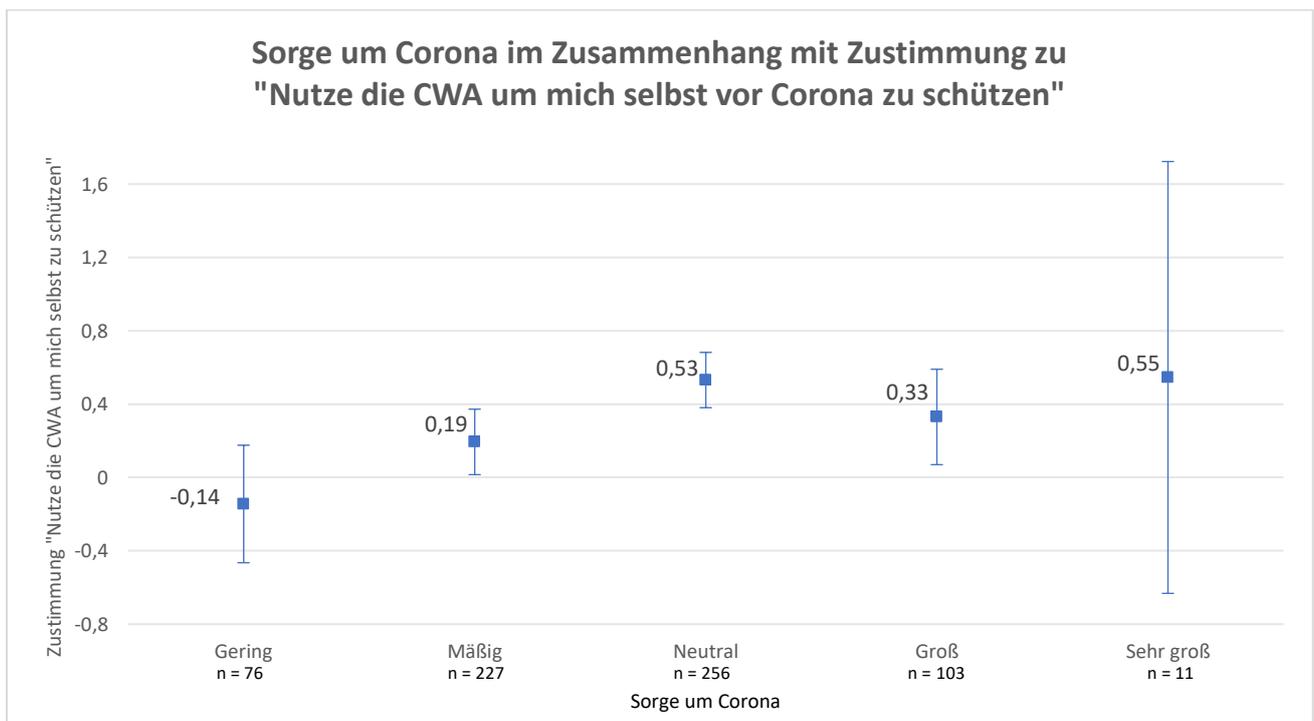

Der Spearman-Rangkorrelationstest ergibt *ρ = 0,12 mit p < 0,005*. Signifikante Unterschiede sind vorhanden zwischen den Gruppen 1 (Geringe Sorge) und 3 (Neutral) mit *p < 0,005* nach paarweisen Vergleich nach Conover mit Bonferroni-Korrektur.

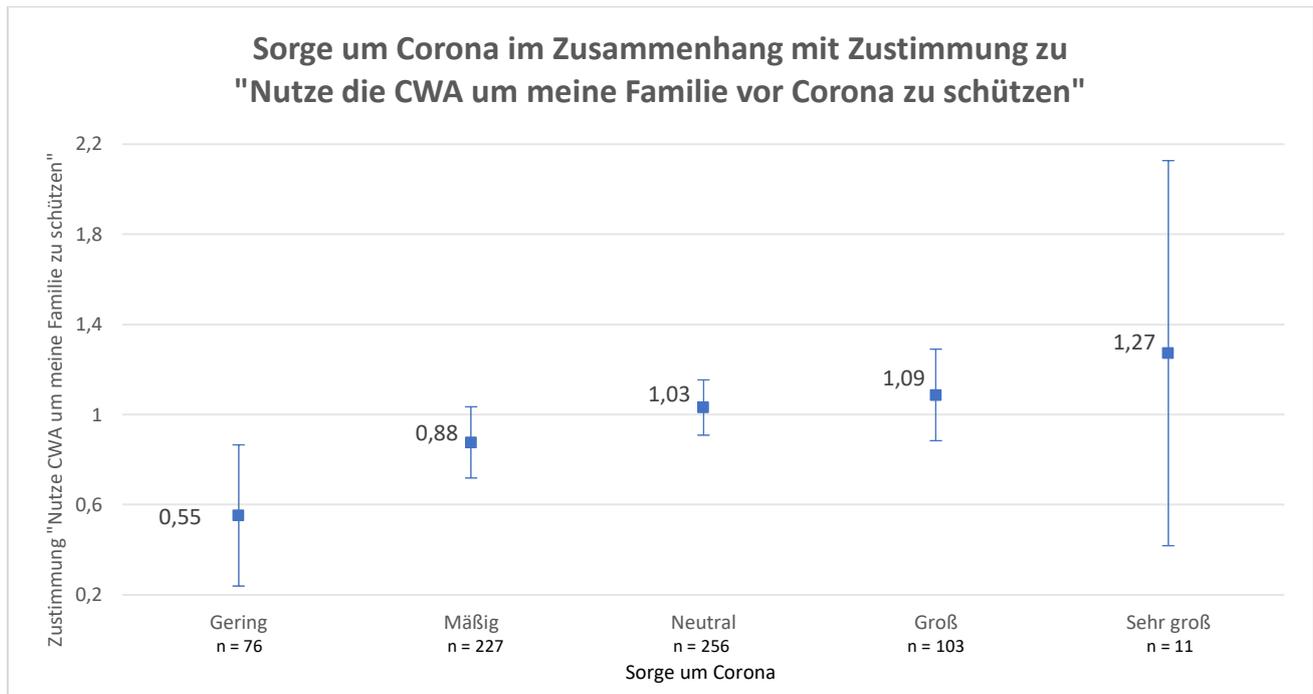

Spearman-Rangkorrelation mit *ρ = 0,11 und p = 0,005*. Keine signifikanten Unterschiede zwischen den Gruppen. Auch bei Personen, die die App häufiger öffnen, sind der Schutz der Familie, der Selbstschutz, und das Beenden der Pandemie als Gründe für die Nutzung der App (noch) stärker ausgeprägt als im Durchschnitt.

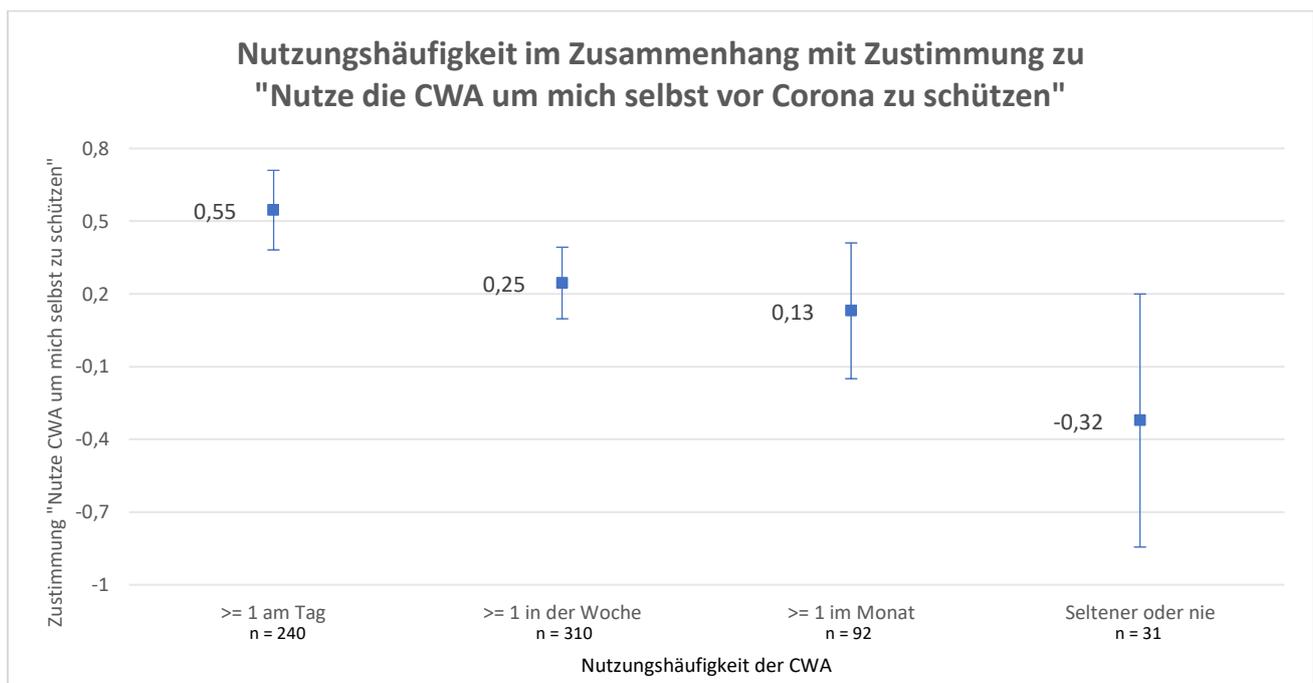

Spearman-Rangkorrelation mit *ρ = -0,15 und p < 0,0005*. Signifikante Unterschiede zwischen den Gruppen {2*,3,4**} und 1 nach Conover-Test mit Bonferroni-Korrektur.

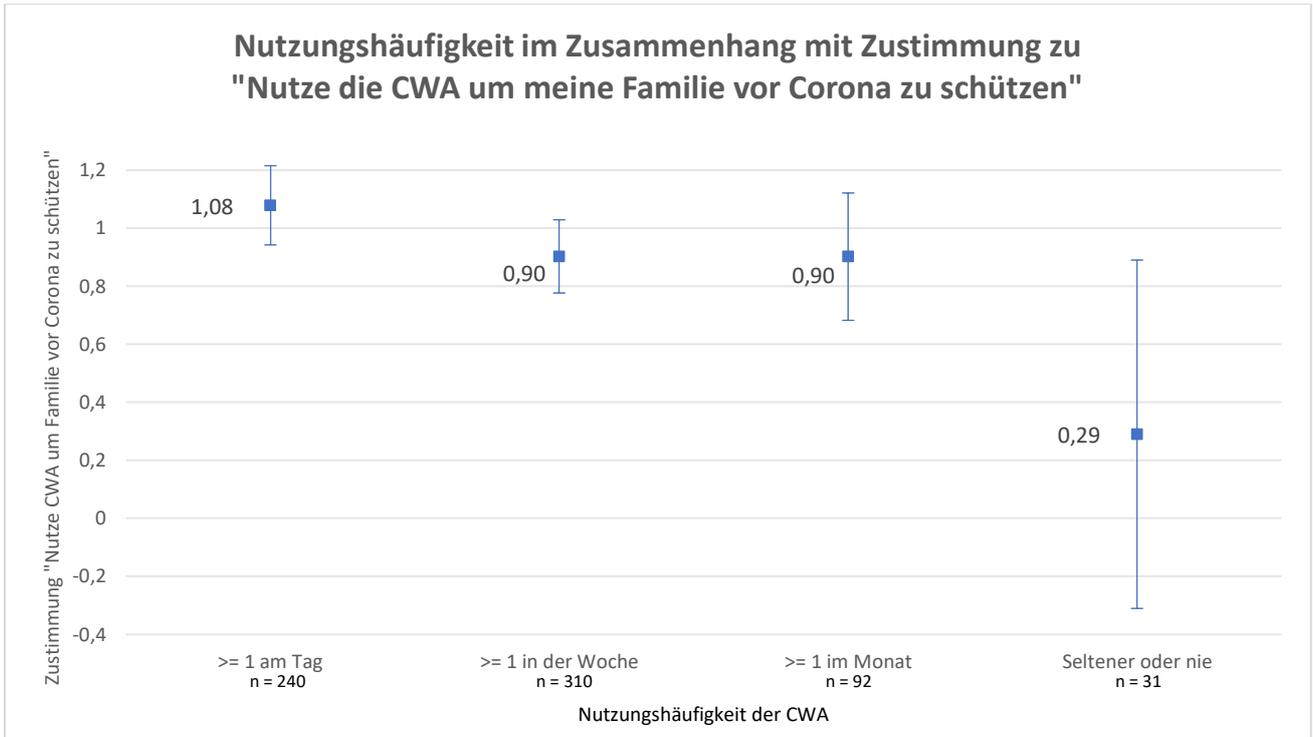

Spearman-Rangkorrelation mit *ρ = -0,11 und p = 0,005*. Es gibt hier keine signifikanten Unterschiede zwischen den Gruppen.

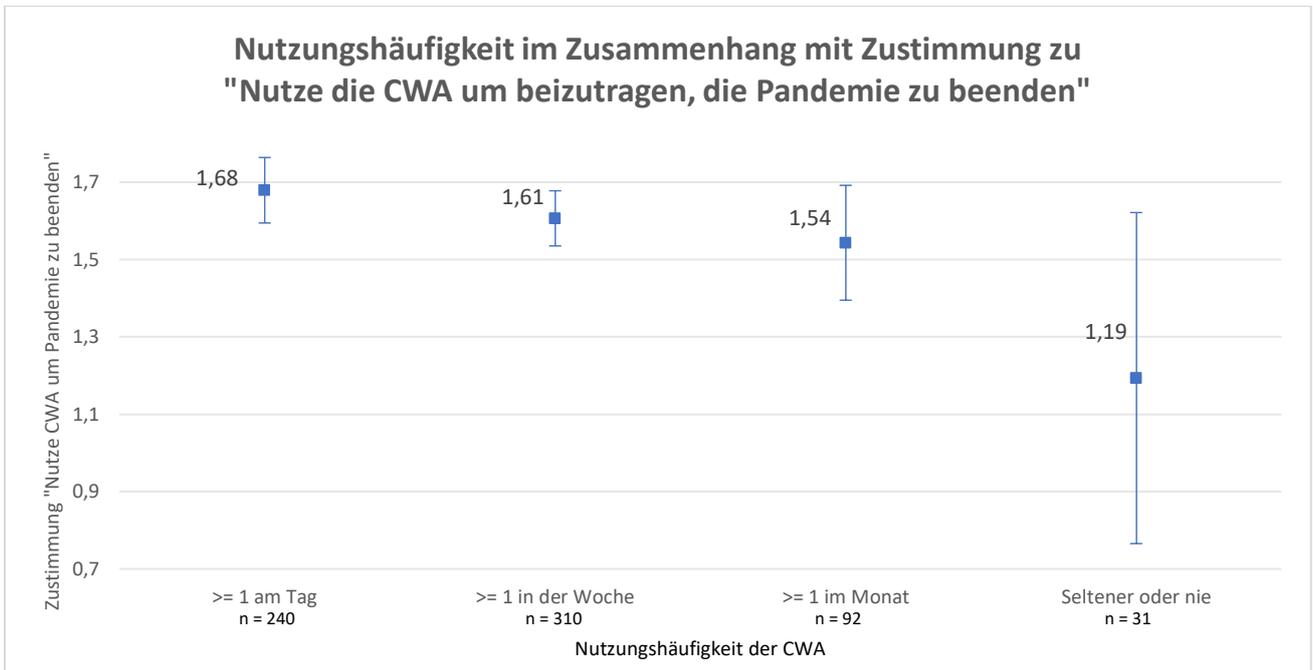

Spearman-Rangkorrelation mit *ρ = -0,12 und p < 0,005*. Es gibt einen signifikanten Unterschied zwischen den Gruppen 1 und 4 mit *p < 0,05* auf Basis eines Conover-Tests mit Bonferroni-Korrektur. Der Grund „um mich selber vor Corona zu schützen" findet in der Altersgruppe 25-29 eine statistisch signifikante leichte Ablehnung.

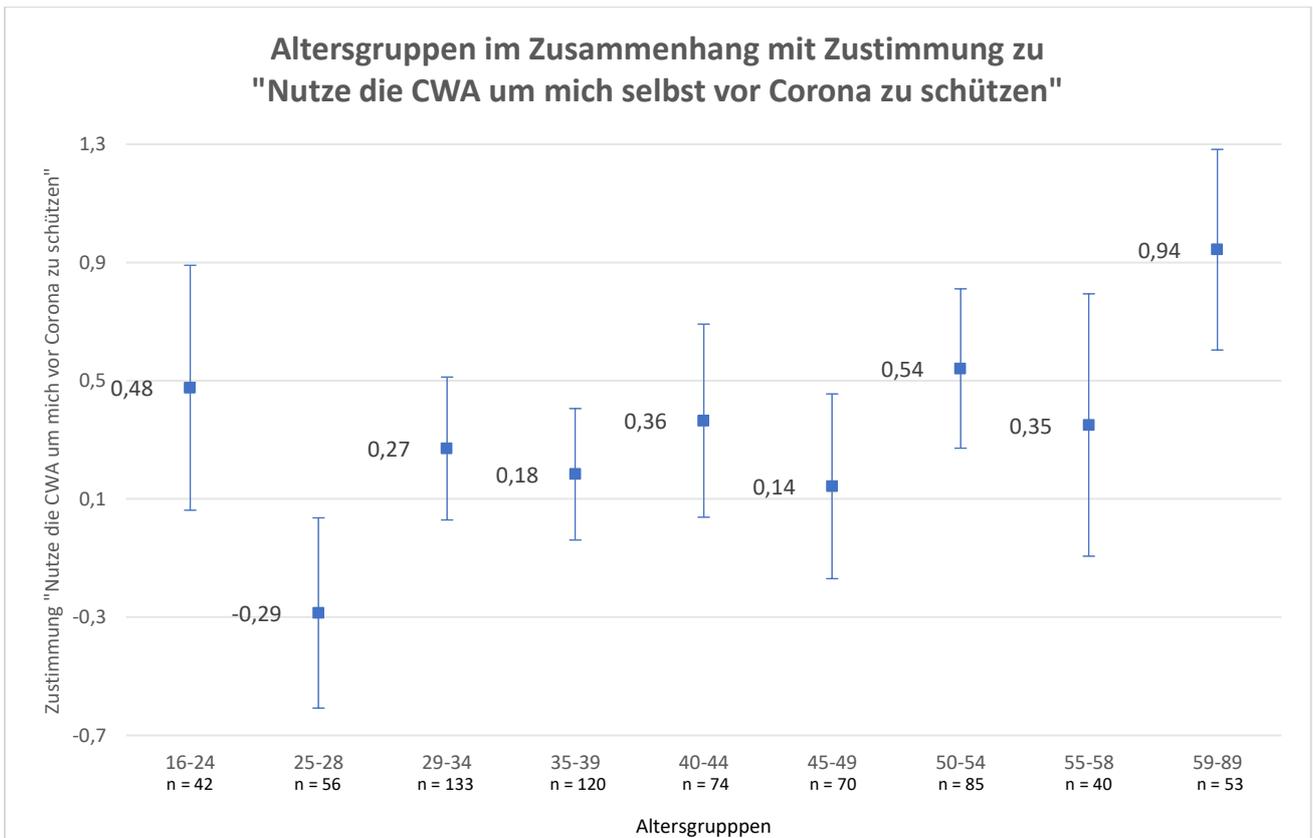

Spearman-Rangkorrelation mit *ρ = 0,13 und p < 0,005*. Es gibt signifikante Unterschiede zwischen den Gruppen {7**} und {2}, zudem zwischen den Gruppen {2***,3*,4**,6*} und {9} auf Basis eines Conover-Tests mit Bonferroni-Korrektur.

## 4.4 Datenschutzverständnis

Das Verständnis bei den Nutzenden zum Datenschutz in der Corona-Warn-App ist hoch. Auf die Fragen, ob sie verstehen, welche Daten erhoben und welche ausgetauscht werden, gab es eine sehr klare Zustimmung. Auch das Vertrauen in den Datenschutz der App und das Verstehen der Risikomessung sind insgesamt sehr deutlich. Nur leicht positiv ist dagegen das Gefühl, ausreichend Kontrolle über die Daten zu haben.

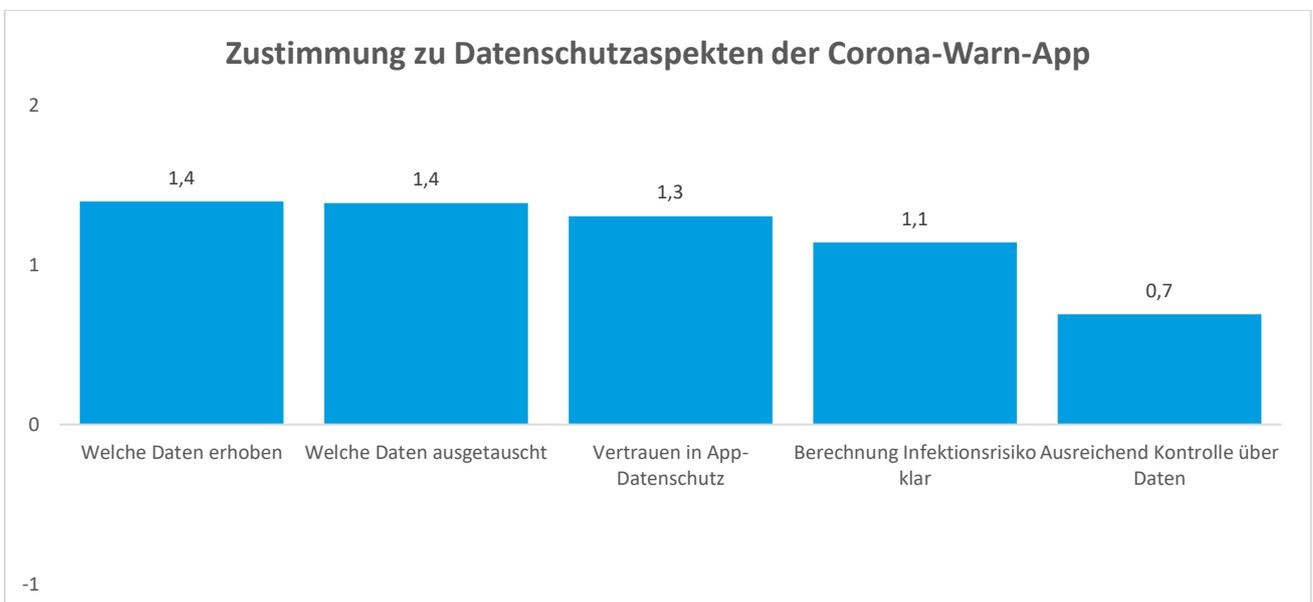

In allen Fällen ist bei Personen mit höherer Technikaffinität die Zustimmung noch stärker ausgeprägt. Auch Personen, denen der Datenschutz wichtig ist, verstehen besser, welche Daten erhoben werden, haben aber kein signifikant anderes Vertrauen in den Datenschutz der App.

Hierbei ist zu beachten, dass die Ausprägung „Geringe Technikaffinität" aus der Analyse ausgeschlossen wurde, da dieser lediglich den Aussagen von 7 Personen zugrunde lag und dies im Vergleich zu den anderen Angaben zu einer sehr hohen Varianz in den Ergebnissen führte und somit nicht sichergestellt werden kann, dass dieses Sample aussagekräftig genug ist.

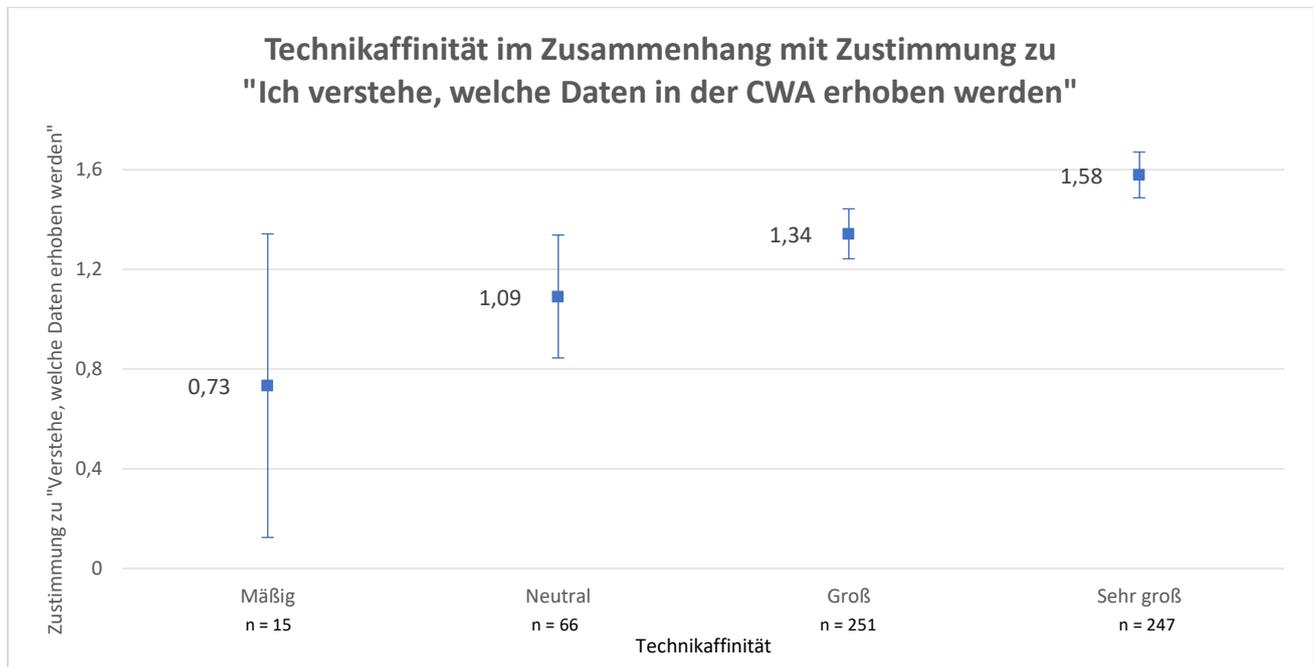

Spearman *ρ = 0,24 mit p = 5,07e-09*. Signifikante Unterschiede zwischen den Gruppen {2\*\*,3\*\*\*,4\*\*\*} und {5} auf Basis eines paarweisen Vergleichs nach Conover mit Bonferroni-Korrektur.

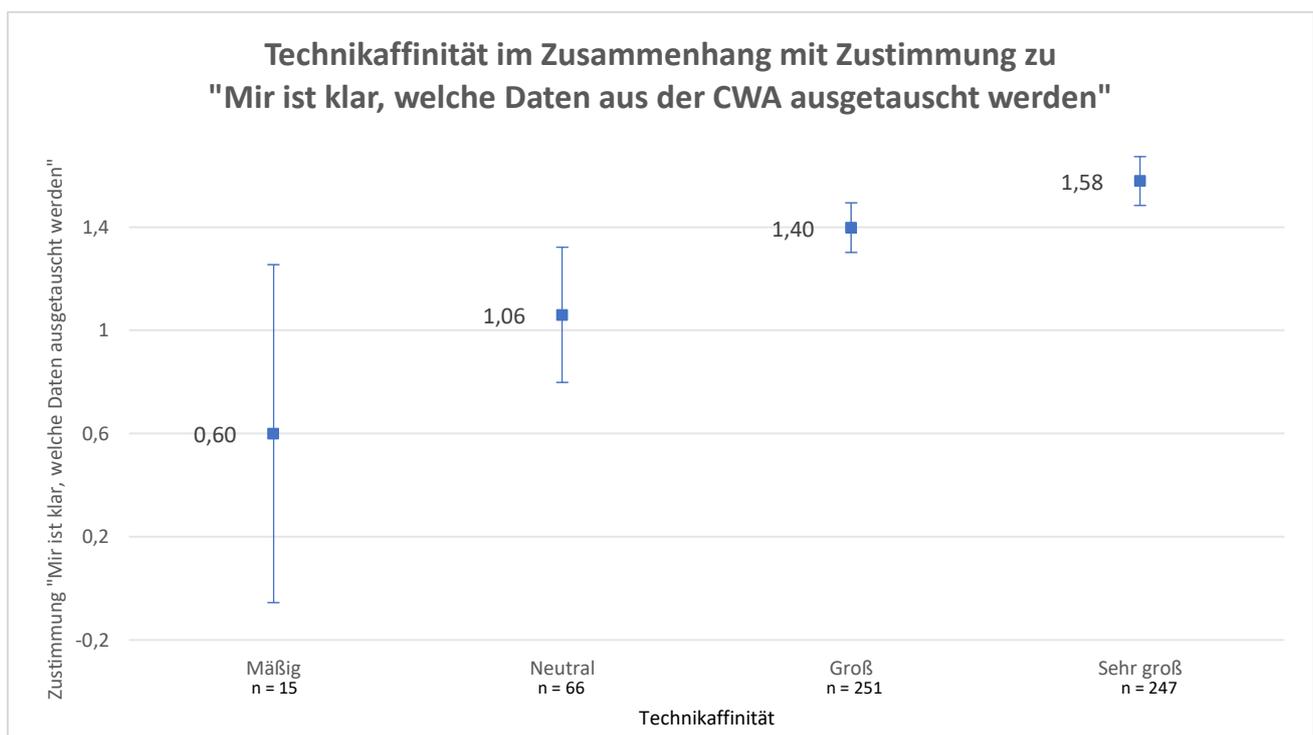

Spearman *ρ = 0,23 mit p = 2,19e-08*. Signifikaten Unterschiede zwischen den Gruppen {2***,3***,4**} und {5} und zwischen {4*} und {2} auf Basis eines paarweisen Vergleichs nach Conover mit Bonferroni-Korrektur.

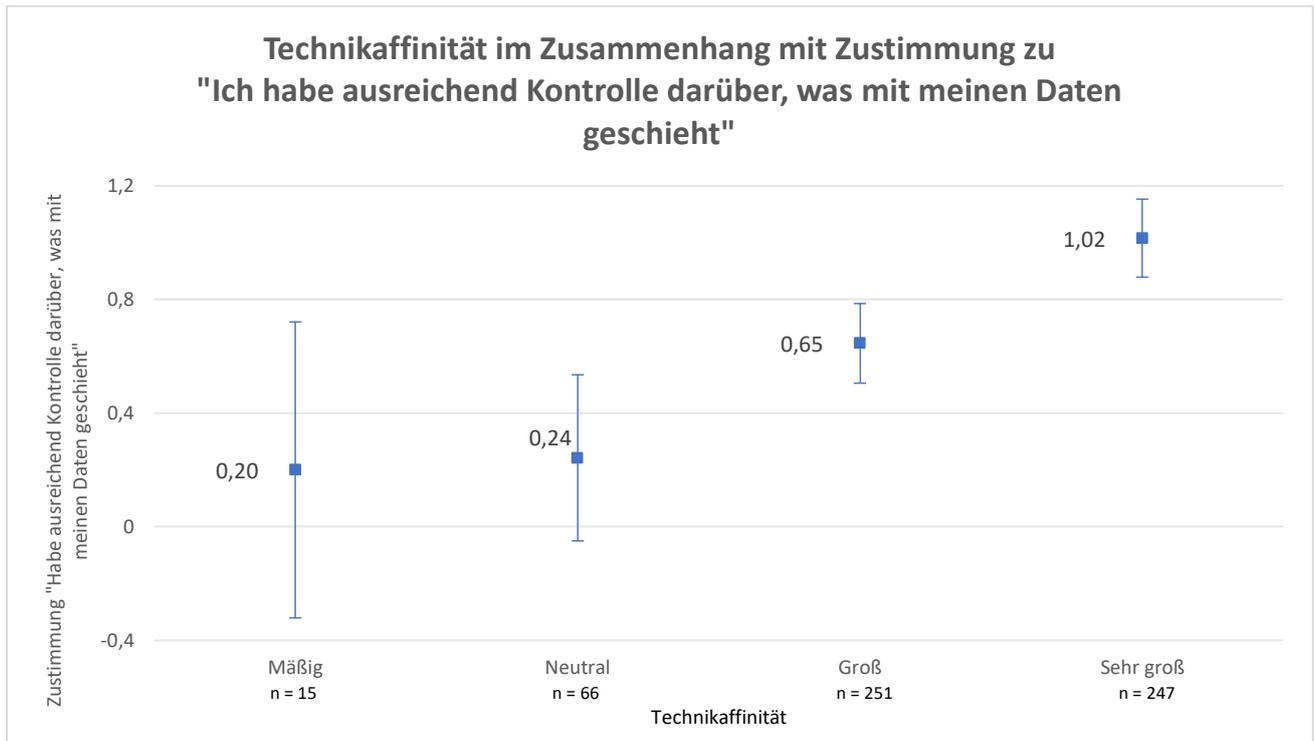

Spearman *ρ = 0,24 mit p = 3,4e-09*. Signifikate Unterschiede zwischen den Gruppen {2*,3***,4**} und {5} auf Basis eines paarweisen Vergleichs nach Conover mit Bonferroni-Korrektur.

### 4.5 Einfluss der Nutzung der CWA auf die Einstellung zu Corona

Der Einfluss der App-Nutzung auf das Verhalten ist eher neutral. Die Fragen, ob sie ein besseres Gefühl von Sicherheit haben, sich im öffentlichen Raum sicherer fühlen, oder sich auf riskante Situationen besser vorbereitet fühlen, wurden jeweils knapp ablehnend beantwortet (-0,05; -0,1; -0,32). Deutlicher ist mit -0,71 die Ablehnung bei der Frage, ob die Nutzenden denken, mit der App eher gesund zu bleiben.

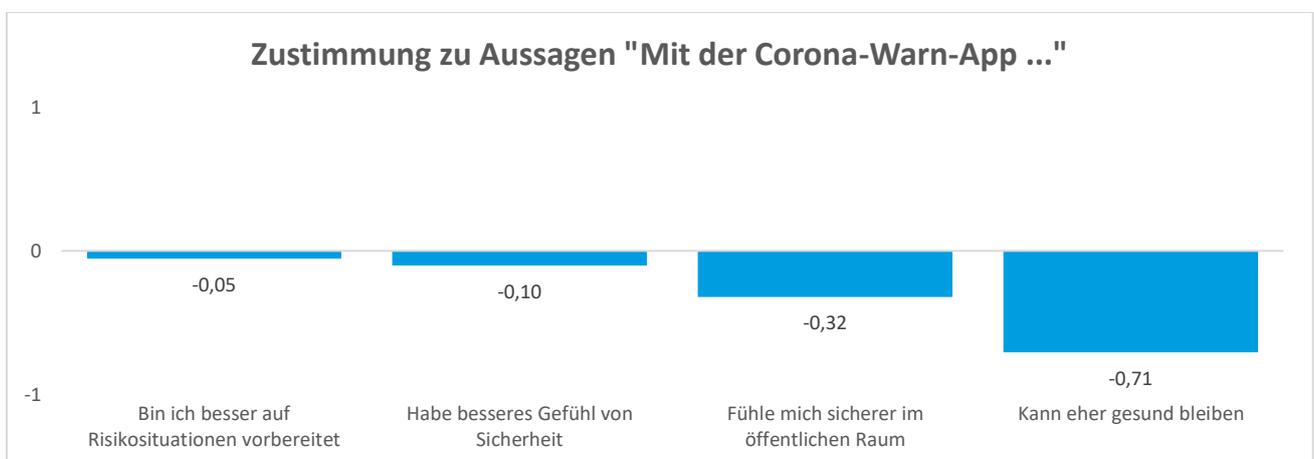

Personen, die die App häufiger öffnen, stimmen den Aussagen eher zu, bzw. haben zumindest hinsichtlich des Gesundbleibens eine weniger starke Ablehnung.

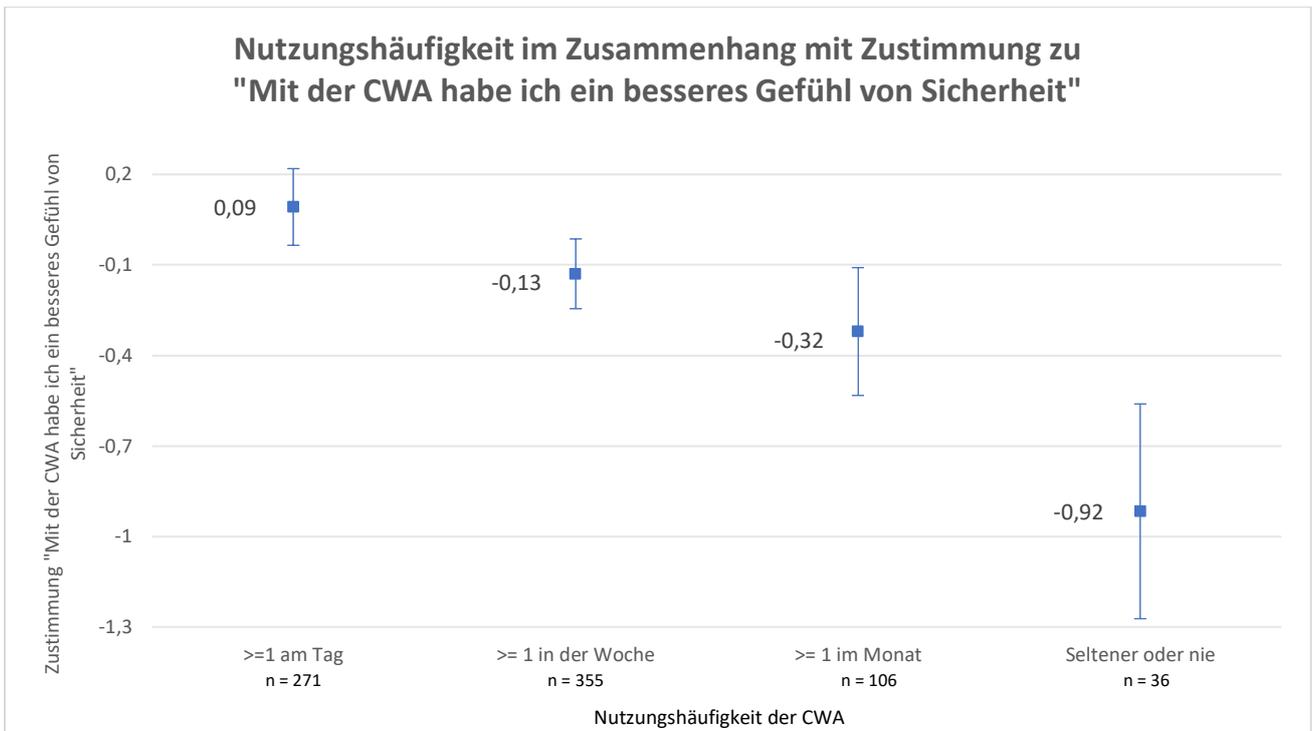

Spearman *ρ = -0,18 mit p = 9,46e-07*. Signifikate Unterschiede zwischen den Gruppen {3**,4***} und {1} und den Gruppen {2***,3*} und {4} auf Basis eines paarweisen Vergleichs nach Conover mit Bonferroni-Korrektur.

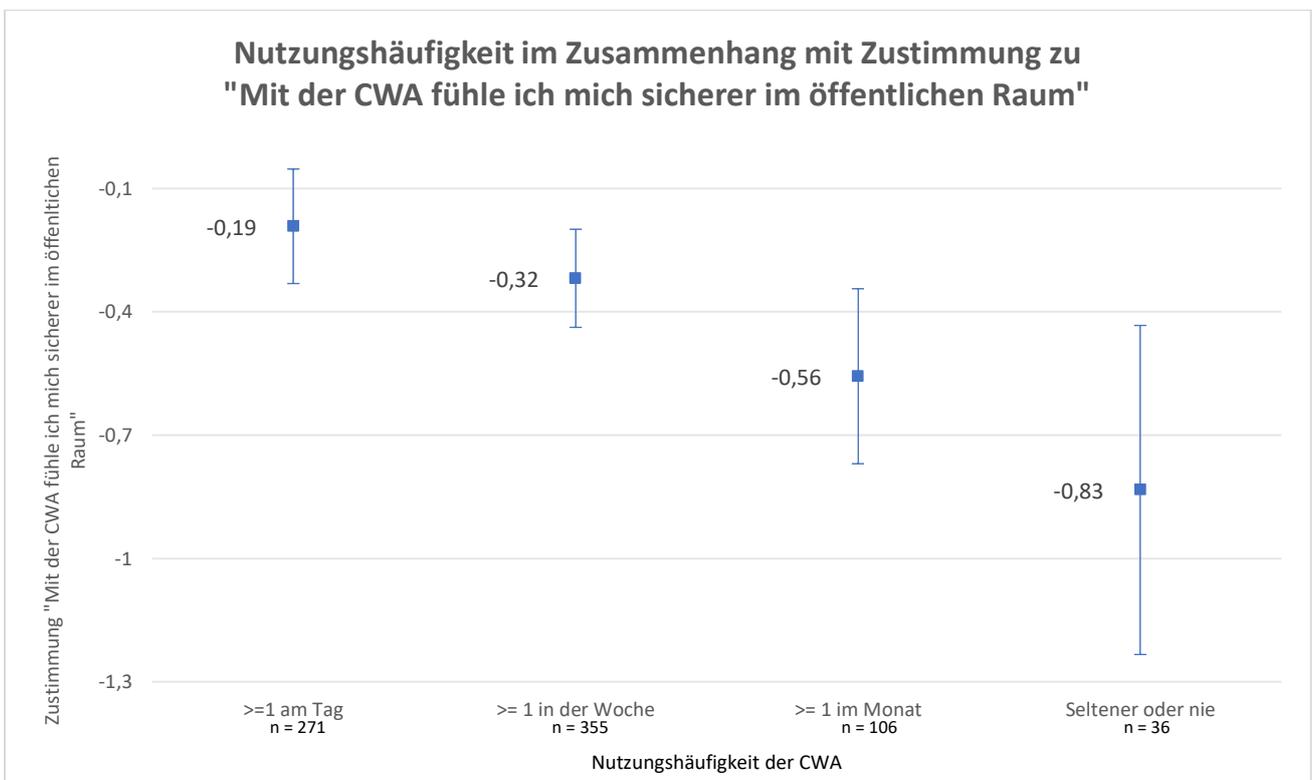

Spearman *ρ = -0,13 mit p < 0,0005*. Signifikate Unterschiede zwischen den Gruppen {3*,4**} und {1} auf Basis eines paarweisen Vergleichs nach Conover mit Bonferroni-Korrektur. Die Tendenz, dass mit Abnahme der Nutzungshäufigkeit der App auch ablehnender in Bezug auf die generelle Einstellung zur Pandemie geantwortet wird, setzt ich auch bei den zwei weiteren Statements „Mit der CWA fühle ich mich auf Situationen mit möglicherweise höherem Corona-Risiko besser vorbereitet" und „Mit der CWA denke ich, dass ich eher gesund bleiben kann", fort. Die Korrelationen nach Spearman sind entsprechend $ρ = \{-0,12;-0,12\}$ und $p < 0,005$ respektive.

Die Ablehnung zur Aussage „Mit der CWA denke ich, dass ich eher gesund bleiben kann" ist in der Altersgruppe der 25 bis 49-jährigen noch deutlicher ausgeprägt.

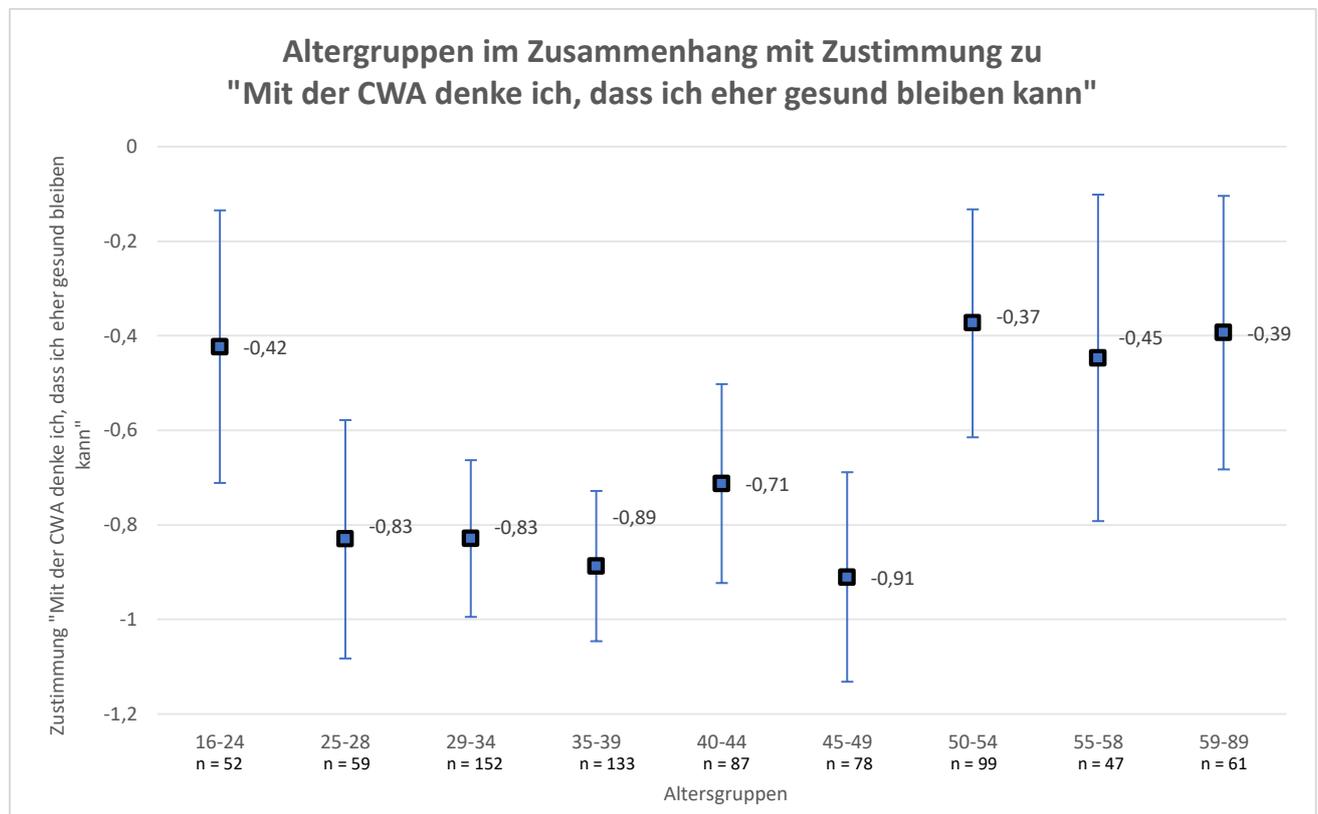

# 5 Vergleich zum infas 360 Sample

Im Folgenden wird das OFFIS-Sample mit dem größeren Sample von infas 360 verglichen. Damit soll deutlich gemacht werden, inwiefern sich die Erkenntnisse aus der vorliegenden Studie mit einer größeren Grundgesamtheit vergleichen lassen. Die infas 360-Umfrage umfasste einen größeren Fragenkatalog als die OFFIS-Studie, zudem wurden auch die demografischen Angaben in einer anderen Form erhoben (z.B. wurde nach Schul- und Berufsbildung unterschieden, im OFFIS-Sample war das nicht der Fall. Dennoch wurde das in 2.2 erwähnte ausgewählte Fragenset aus der OFFIS-Umfrage 1:1 in die infas 360-Umfrage übernommen. Diese gemeinsamen Fragen werden im Folgenden miteinander verglichen, um zu überprüfen, inwiefern sich die OFFIS- und infas 360 Samples unterscheiden. Der Grund für die hohe Zahl an Teilnehmenden der infas 360-Studie begründet sich mit der Regionalisierung der Befragungsergebnisse[3].

Demografisch gesehen liegt das Durchschnittsalter bei 41,3 Jahren ($SD = 11,9$) im OFFIS-Sample und bei 47,7 Jahren ($SD = 16,4$) beim infas 360 Sample. Die Frauenquote liegt beim infas 360 Sample bei 49 %, beim OFFIS-Sample bei 60 %. Die Verteilung des Bildungsniveaus ist in der folgenden Grafik abgebildet. Beim OFFIS-Sample ist eine starke Tendenz zugunsten der studierten und promovierten zu erkennen. Hierbei zu beachten, dass dieser Vergleich entlang der OFFIS-Kategorien vorgenommen wurde, da das infas 360 Sample zwischen Schulbildung und Berufsbildung (u.a. Studium und Promotion) unterschied und auch noch weitere Schul- und Berufsausbildungskategorien erfasste.

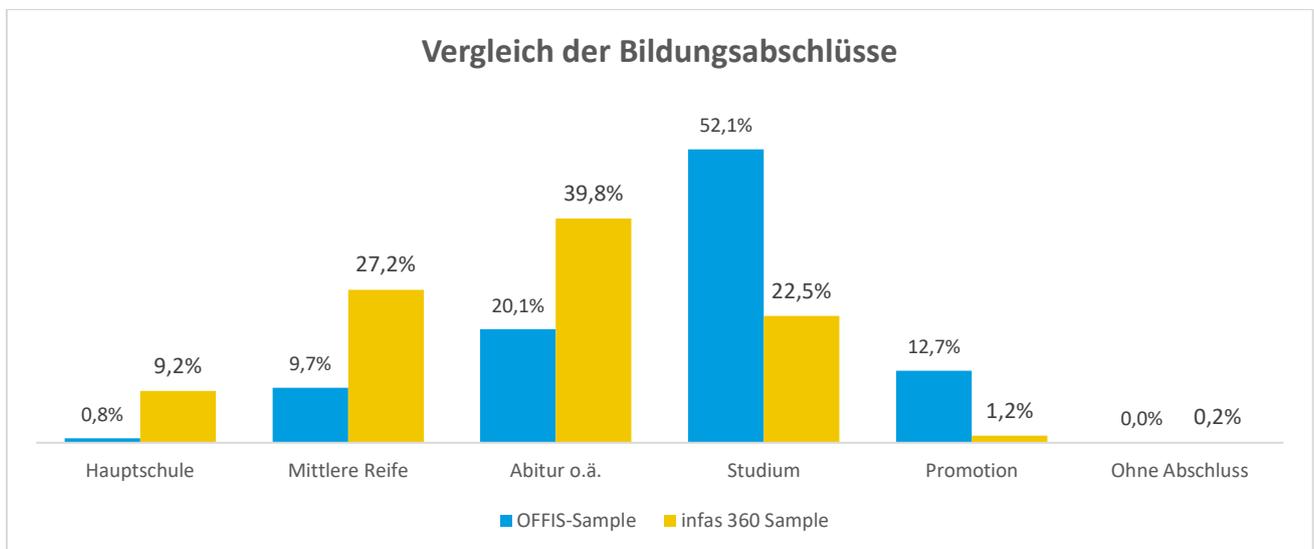

Im Hinblick auf die allgemeine Nutzungsgewohnheiten ist festzustellen, dass in Bezug auf die tägliche Mitnahme des eigenen Handys keine großen Unterschiede bestehen. Dazu gibt es im OFFIS-Sample die Möglichkeit „keine Angabe" zu machen. Dies war bei der infas 360-Umfrage nicht der Fall. Unter Umständen könnten sich deswegen die Werte noch weiter angleichen.

---

[3] Rückfragen zum infas 360-Sample können direkt bei infas 360 an Frau Kroth (j.kroth@infas360.de) gerichtet werden.

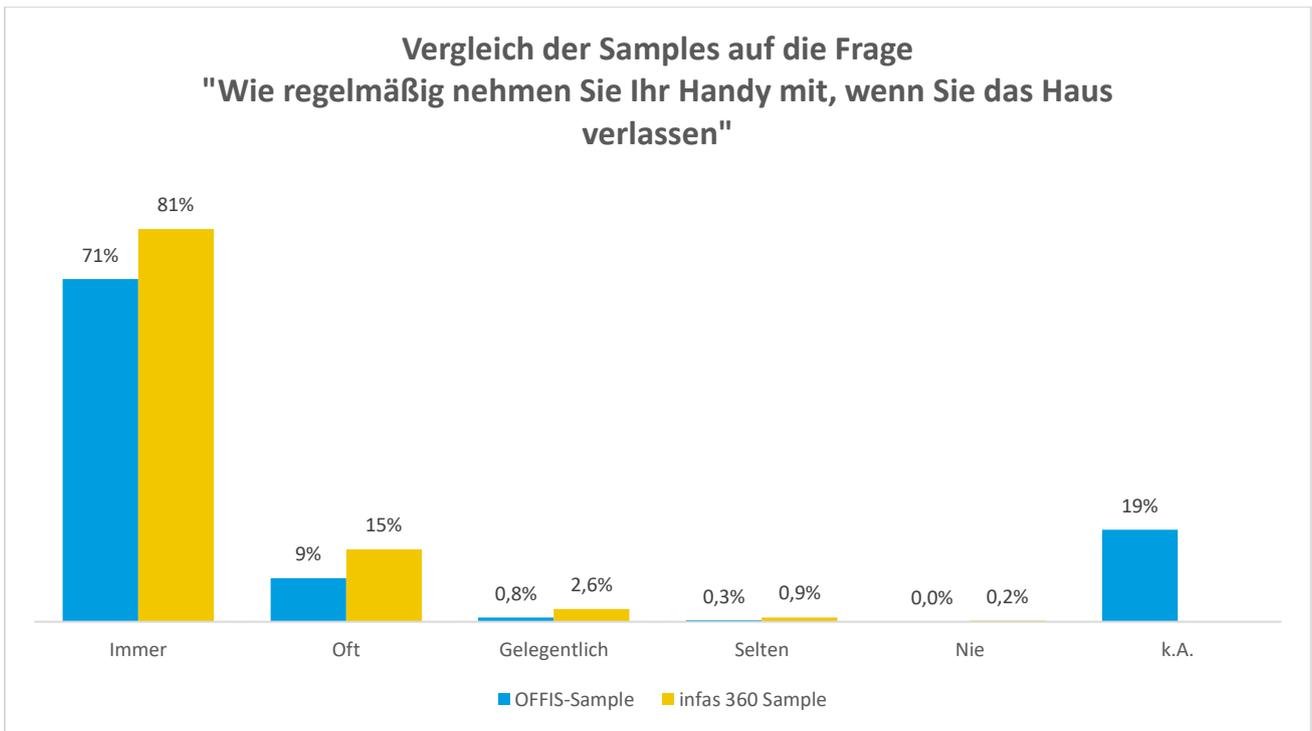

Bei Betrachtung der Zahlen zur aktiven Nutzung der App fallen auch hier keine strukturellen Unterschiede auf. Beim OFFIS-Sample sind alle Werte max. 9 Punkte unter dem infas 360 Sample, was sich auch hier evtl. mit der recht hohen Zahl von „keine Angabe" von 19 % im OFFIS-Sample begründen ließe.

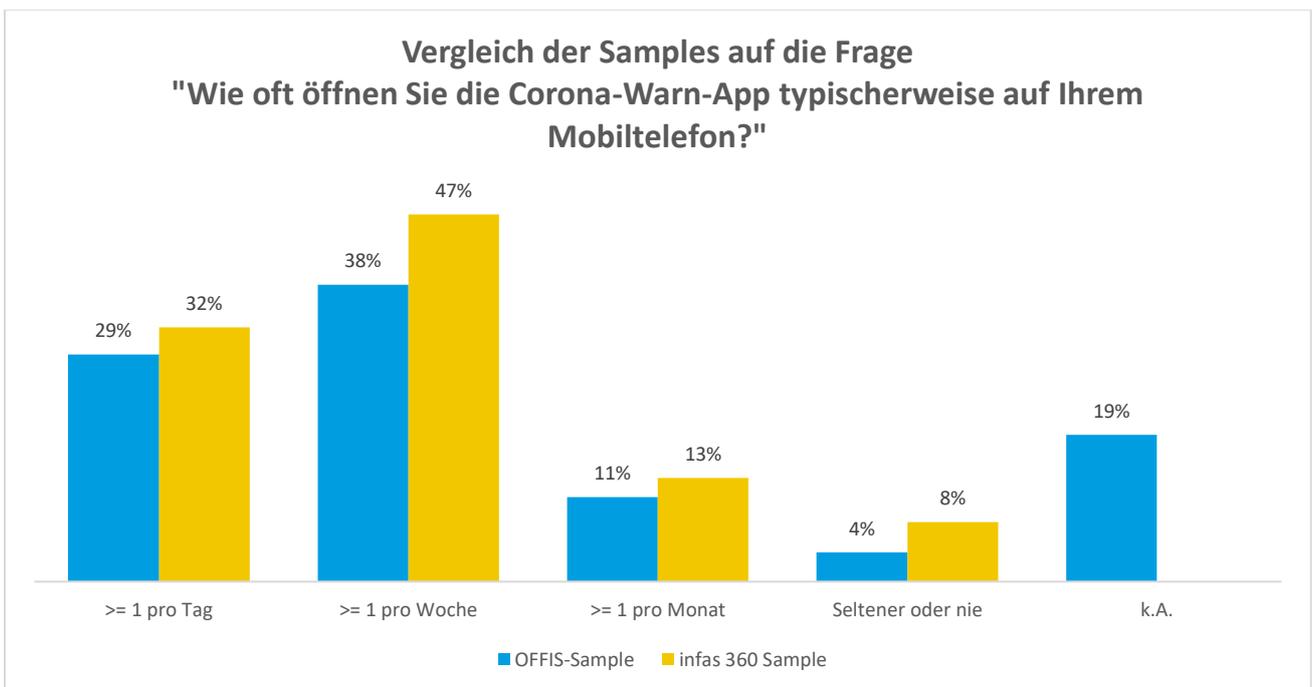

## 5.1 Nicht-Nutzung

Der Anteil der Nicht-Nutzenden insgesamt ist im infas 360-Sample erheblich höher (63 %) als im OFFIS-Sample (19 %). Der wesentliche Unterschied liegt bei denjenigen, die die App nie installiert haben (OFFIS: 11 %, infas 360: 53 % aller Befragten), während die Unterschiede bei denjenigen, die die App erfolglos versucht haben zu installieren, sie nach Installation nicht oder nicht mehr aktiviert haben, oder wieder deinstalliert haben, in beiden Samples gering sind.

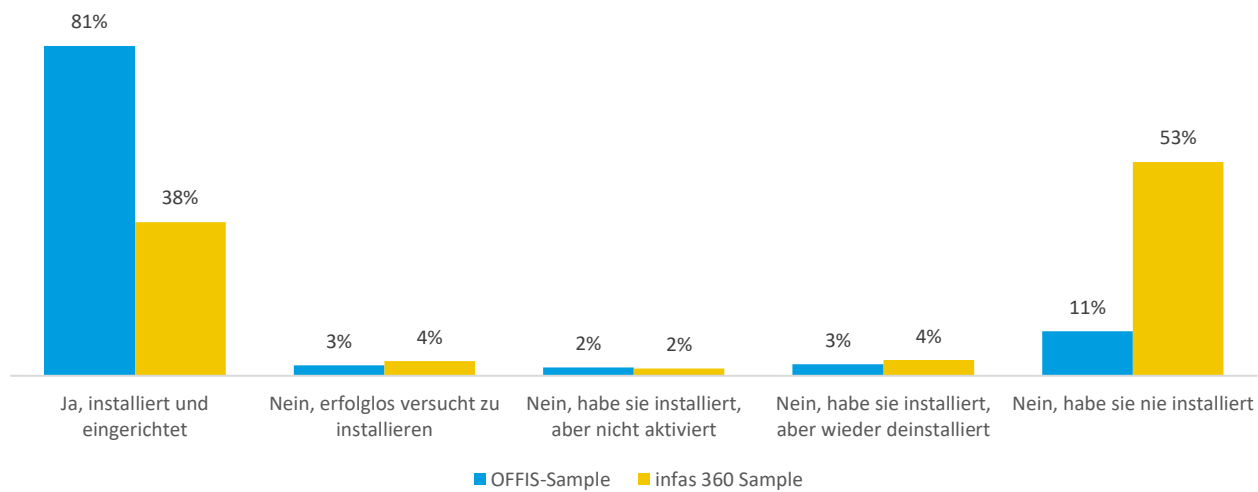

Bei den Gründen für die Nicht-Nutzung zeigt sich – trotz demografischer Unterschiede zwischen den beiden Samples – ebenfalls eine sehr ähnliche Verteilung im OFFIS und im infas 360 Sample.

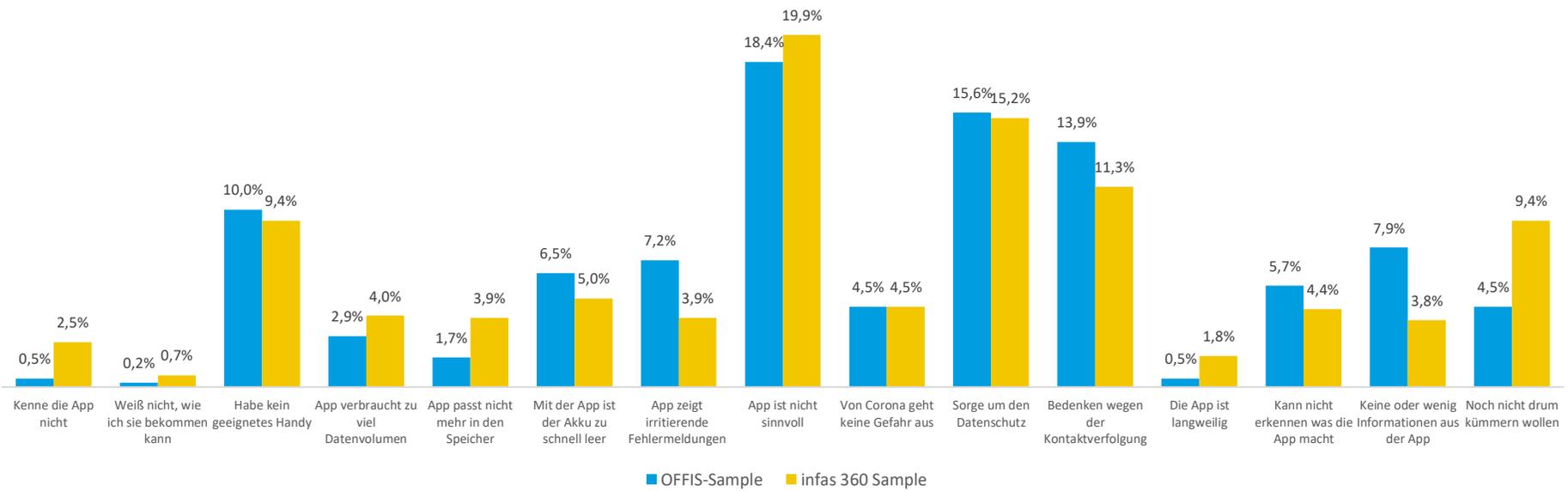

## 5.2 Art der Nutzung und Gründe für die Nutzung

Die Verteilung der Gründe für die Nutzung der CWA deckt sich ebenfalls weitgehend mit der des OFFIS-Samples. Der (objektiv nicht sinnvolle) Grund „um mich selbst vor Corona zu schützen" findet jedoch im infas 360 Sample mehr Zustimmung (OFFIS: 0,30; infas 360: 1,3). Die Unsicherheit, ob das Umfeld oder die Werbung zur Nutzungsentscheidung beigetragen haben, ist – bei insgesamt übereinstimmender eher Ablehnung – im OFFIS-Sample größer. (Die Frage, ob Werbung dazu beigetragen hat, wurde im infas 360-Sample nicht gestellt.)

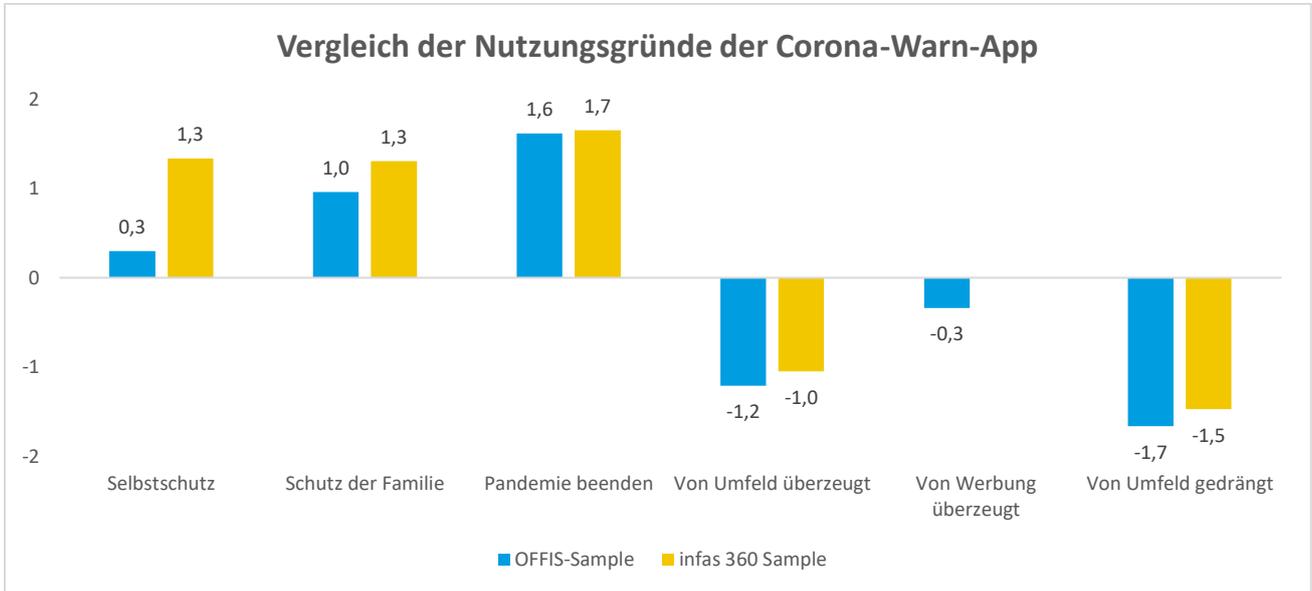

Bei den Gründen für das aktive Öffnen der Corona-Warn-App zeichnet sich ein etwas anderes Bild. Hier sind lediglich zwei Antworten ähnlich verteilt (Corona-Status überprüfen und Funktion der App sicherstellen).

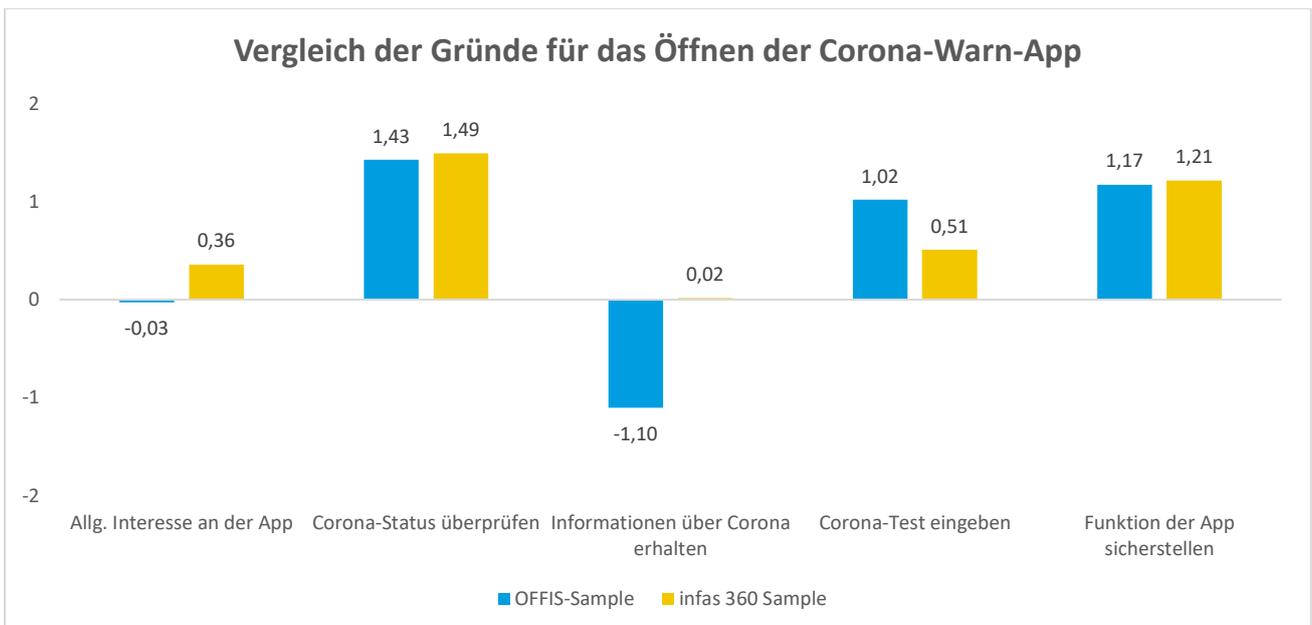

Bei den anderen Antworten gibt es recht deutliche Unterschiede, insbesondere beim Punkt „Informationen über Corona erhalten" und „Corona-Test eingeben". Zudem ist auch das allgemeine Interesse an der App leicht unterschiedlich verteilt.

# 6 Diskussion
## 6.1 Zusammenfassung und Bewertung

Die Ergebnisse dieser Studie basieren auf insgesamt 11879 Antworten in zwei Samples. Das OFFIS-Sample mit einem *n=1326* ist stärker weiblich, hat eine leicht verschobene Altersverteilung und ein deutliches Bias zu höheren Bildungsabschlüssen. Im Vergleich dazu ist das infas 360-Sample mit *n=10553* näher am allgemeinen Bundesschnitt, jedoch nicht repräsentativ. Das infas 360-Sample enthält eine Teilmenge der Fragen des OFFIS-Samples, wie in Abschnitt 2.2 beschreiben. Trotz der Unterschiede zwischen den beiden Samples zeigen sich gemeinsame Tendenzen, wie und warum die Corona-Warn-App genutzt und nicht genutzt wird.

Auf Grundlage unserer Daten ergeben sich **vier Hauptgründe** für das Nicht-Nutzen der Corona-Warn-App. Dabei stehen an erster Stelle **Datenschutzbedenken**. Nahezu gleichauf sind **Zweifel am Sinn der App** sowie **technische Probleme**. Schließlich werden auch Probleme genannt, die sich allgemein unter User Experience zusammenfassen lassen, und die widerspiegeln, dass die Nutzenden den Umgang mit der App nicht angenehm finden. Weniger relevant, aber sichtbar waren Gleichgültigkeit sowie grundsätzliche Zweifel an der Gefahr durch Corona. Es ist bemerkenswert, dass gerade Datenschutzbedenken an erster Stelle stehen, obwohl Datenschützer und IT-Experten sehr einhellig den Datenschutz der CWA loben. Hinsichtlich der Zweifel an der App bleibt unklar, ob es Zweifel am korrekten Funktionieren der App zum Erkennen von Risikosituationen ist, ob allgemein die Funktion der App nicht verstanden wird, oder ob andere Gründe dahinterstehen.

Bei denjenigen, die die Corona-Warn-App nutzen, ist sie in der Regel aktuell und das Handy wird meistens oder immer mitgeführt, so dass das **grundsätzliche Funktionieren im Wesentlichen sichergestellt scheint**.

Die Nutzenden **öffnen die App regelmäßig**, oft mindestens wöchentlich. Hauptgründe sind das **Überprüfen des eigenen Status** und das **Sicherstellen des Funktionierens der App**. Letzteres ist technisch nicht oder nicht mehr notwendig und wurde zum Zeitpunkt der Umfrage nicht ausdrücklich empfohlen. Mittlerweile empfangen Nutzende in regelmäßiger Weise Benachrichtigungen von der App, den eigenen Risikostatus zu überprüfen. Das regelmäßige Öffnen und die Funktionsüberprüfung ist bei Personen **mit höherer Corona-Sorge stärker** ausgeprägt und **nimmt insgesamt zu, wenn in räumlicher oder sozialer Nähe das Risiko steigt**.

Auffallend ist, dass ein deutlicher Anteil der Nutzenden **unsicher war, ob sie die App zum Eingeben von Testergebnissen öffnen** würden. Unklar bleibt hier, ob lediglich eine Unkenntnis darüber herrscht, ob und wie Testergebnisse in der App eingegeben werden, oder ob es ein Misstrauen darüber ist, ein Testergebnis eingeben zu wollen. Die Zustimmung war zudem im infas 360-Sample noch schwächer ausgeprägt und ist mit 0,51 sehr nah am Neutralniveau.

Die App wird nur von **wenigen Nutzenden zum Abrufen von Corona-Informationen verwendet**, etwas stärker von Älteren in der Altersgruppe 50+, wobei angemerkt werden muss, dass der Informationsaspekt keine Kernfunktion der App darstellt.

Die **Gründe für die Nutzung der App** sind vor allen Dingen altruistisch: Die Nutzenden wollen einen **Beitrag dazu liefern, die Pandemie zu beenden**. Nachgeordnet, aber noch deutlich ist der Grund, das **eigene Umfeld zu schützen**. Deutlich geringer, aber ebenfalls noch zustimmend ist der **Selbstschutz**, obwohl dies objektiv keine Funktion der App ist. Die Wirkweise der App wird möglicherweise – auch wenn sie vielen im Grundsatz klar ist – von einem deutlichen Anteil der Nutzenden nicht vollständig verstanden. Bemerkenswert ist auch, dass scheinbar, zumindest bis zum Zeitpunkt der Umfrage, Werbung und Nachrichten nur wenig dazu beigetragen haben, die App zu nutzen. Das ist zumindest teilweise dadurch zu erklären, dass die meisten Nutzenden die App bereits sehr früh installiert haben, als sie noch wenig präsent in den Medien war.

Die Nutzenden geben an, den **Datenschutz insgesamt gut zu verstehen**. Lediglich bei der **empfundenen Kontrolle über die eigenen Daten gibt es eine leichte Zurückhaltung**. Dieses hohe, wenn auch möglicherweise nur gefühlte, Datenschutzwissen ist ein bemerkenswerter Kontrast dazu, dass für die Nicht-Nutzenden Datenschutzbedenken der Haupt-Ablehnungsgrund waren. Es bleibt unklar, ob das Verstehen des Datenschutzkonzeptes der App die Installationsbereitschaft positiv beeinflusst hat, und ob die Konzepte tatsächlich verstanden werden. Die Zurückhaltung beim Kontrollempfinden deutet zumindest darauf hin, dass auch bei den Nutzenden ein restliches Unwohlsein hinsichtlich des Datenschutzes verbleibt.

**Neutral ausgeprägt bleibt die Beeinflussung des eigenen Sicherheitsempfindens** und des Verhaltens im öffentlichen Raum bezüglich Corona-Risiken. Da die App aber objektiv keine direkte individuelle Sicherheit bietet, wäre eine Ablehnung dieser Beeinflussung angemessener gewesen. Durch die App könnte damit ein zumindest leichter „Airbag-Effekt" hervorgerufen werden, wonach eine höhere empfundene Sicherheit zu einem riskanteren Verhalten führt.

## 6.2 Folgerungen

Aus der Studie lassen sich zwei mögliche, sich ergänzende Folgerungen ziehen. Zum einen lässt sich u.a. aufgrund der Zunahme der App-Öffnungszahlen bei Verschlechterung der Corona-Situation vermuten, dass die Nutzenden mit der Corona-Warn-App interagieren wollen, und sie nicht ausschließlich als rein passive Hintergrund-App auf dem Handy laufen lassen wollen. Zum anderen wird möglicherweise die Funktion der App nicht ausreichend verstanden, sodass es zu Zweifeln an der Sinnhaftigkeit der App und an ihrem Datenschutzkonzept kommt. Die begründet sich mit den allgemeinen Datenschutzbedenken und dem erhöhten Sicherheitsgefühl bei Personen, die die App häufiger öffnen.

Wir schlagen daher drei Maßnahmen vor, um die Akzeptanz der App in der Bevölkerung erhöhen können:

**Maßnahme 1:** Die App sollte stärker interaktive Elemente und Funktionen bereitstellen, die ihre Nutzung attraktiver machen. Das – zum Zeitpunkt der Studiendurchführung noch nicht implementierte – Symptomtagebuch ist eine von vielen denkbaren Elementen.

**Maßnahme 2:** Die internen Vorgänge in der App, die heute sehr weitgehend verborgen ablaufen, sollten stärker transparent gemacht werden, so dass die Nutzenden ein besseres Empfinden darüber bekommen, was die App im Allgemeinen und mit ihren Daten macht.

**Maßnahme 3:** Schließlich ist eine bessere allgemeine Aufklärung darüber notwendig, was die App genau macht und wie sie funktioniert, um so die Zweifel der Nicht-Nutzenden am Datenschutz und am Sinn der App zu beseitigen und damit die Anzahl der Nutzenden zu erhöhen.

Weitere Maßnahmen, insbesondere eine Beseitigung von technischen Barrieren, sind grundsätzlich empfehlenswert.

## 6.3 Einschränkungen der Studie

Die Studie gibt einen ersten Blick in die Art und Weise und Gründe der Nutzung und Nicht-Nutzung der Corona-App, bleibt aber grundsätzlich beschränkt. Das OFFIS-Sample, welches tiefere Einblicke erlaubt, hat einen Bias zu höheren Bildungsabschlüssen. Das infas 360-Sample ist deutlich größer, aber ebenfalls nicht repräsentativ, und betrachtet nur eine Teilmenge der Fragen.

Als recht gut belegt kann man diejenigen Aussagen annehmen, die sich sowohl im OFFIS-Sample als auch im infas 360-Sample finden und die weitgehend vergleichbaren Antworten geben: Gründe für Nichtnutzung, Umgang mit App und Handy, Gründe für die Nutzung der App, Gründe für das Öffnen der App.

Die hohe Übereinstimmung in diesen Aussagen zwischen den beiden Samples ist zwar ein Indiz dafür, dass auch die weiteren Aussagen aus dem OFFIS-Sample (Zusammenhänge zu Datenschutz, Corona-Sorge, Technik-Affinität, Alter, Ausbildungsstand, Verständnis von Datenschutz der CWA, Einfluss der CWA auf das Corona-Verhalten) trotz des starken Bias in Richtung hoher Bildungsabschlüsse tendenziell korrekt sind. Das bleibt aber spekulativ.

Auch die demografischen Fragen und diejenigen zur Corona-Sorge, Technikaffinität und Datenschutzbedeutung sind – dem Design der Studie folgend – sehr knapp ausgefallen und können nicht als endgültig verlässlich angesehen werden, da sie lediglich eine informierende Bedeutung haben und nicht das gesamte Spektrum der Selbsteinschätzung abbilden.

Trotz der Einschränkungen liefert die Studie wichtige Erkenntnisse, die auch im größeren infas 360-Sample bestätigt wurden. Die beschriebenen Einschränkungen sollten in vertiefenden, möglichst repräsentativen Folgeuntersuchungen berücksichtigt werden. Gerade auch bei anstehenden Veränderungen der CWA gilt es, ein qualitativ hochwertiges Monitoring der Nutzung zu sichern.

# 7 Anhang: Vollständige Fragenliste

1. Einstiegsfragen
    1.1. Hiermit bestätige ich, dass ich 16 Jahre oder älter bin.
    1.2. Nutzen Sie die Corona-Warn-App? Antwortmöglichkeiten: Ja, sie ist installiert und eingerichtet. Nein, ich habe erfolglos versucht sie zu installieren. Nein, ich habe sie installiert, aber nicht aktiviert oder später wieder deaktiviert. Nein, ich habe sie installiert, aber später wieder deinstalliert. Nein, ich habe sie nie installiert.
2. Nur falls 1.2 nicht mit „ja" beantwortet wurde: Gründe der Nichtnutzung
    2.1. Warum nutzen Sie die App nicht oder nicht mehr? Antwortmöglichkeiten (Mehrfachantwort möglich)
        2.1.1. Ich kenne die App nicht.
        2.1.2. Ich weiß nicht, wie ich sie bekommen kann.
        2.1.3. Ich habe kein geeignetes Mobiltelefon.
        2.1.4. Die App verbraucht zu viel Datenvolumen.
        2.1.5. Die App passt nicht mehr in den Speicher meines Mobiltelefons.
        2.1.6. Mit der App ist der Akku zu schnell leer.
        2.1.7. Die App zeigt irritierende Fehlermeldungen.
        2.1.8. Ich denke die App ist nicht sinnvoll.
        2.1.9. Ich glaube, von Corona geht keine Gefahr aus.
        2.1.10. Ich habe Sorge um den Datenschutz.
        2.1.11. Ich habe Bedenken wegen der Kontaktverfolgung.
        2.1.12. Die App ist langweilig.
        2.1.13. Ich kann nicht erkennen, was die App macht.
        2.1.14. Ich bekomme keine oder wenig Informationen aus der App.
        2.1.15. Ich habe mich bisher noch nicht darum kümmern wollen.
        2.1.16. Sonstiges: (Freitext)
3. Nur falls 1.2 mit „ja" beantwortet wurde: Nutzungsdauer und –Häufigkeit
    3.1. Wie lange nutzen Sie die Corona-Warn-App bereits? Antwortoptionen: Seit bis zu 1 Monat. Seit bis zu 2 Monate. Länger als 2 Monate. Seit der Veröffentlichung Mitte Juni 2020
    3.2. Achten Sie darauf, dass Ihre Corona-Warn-App aktualisiert wird? Antwortoptionen: Ja, Nein, Weiß ich nicht.

- 3.3. Wie regelmäßig nehmen Sie Ihr Handy mit, wenn Sie das Haus verlassen? Antwortoptionen: Nie, Selten, Gelegentlich, Oft, Immer.
- 3.4. Wie oft öffnen Sie die Corona-Warn-App typischerweise auf Ihrem Mobiltelefon? Antwortoptionen: Ein oder mehrmals pro Tag, Ein oder mehrmals pro Woche, Ein oder mehrmals pro Monat, Seltener oder nie
- 3.5. Öffnen Sie die Corona-Warn-App häufiger, wenn sich die äußeren Umstände ändern? Ich denke, dass ich die Corona-Warn-App häufiger öffne, wenn… (Antwortoptionen: Stimme überhaupt nicht zu | Stimme eher nicht zu | Neutral | Stimme eher zu | Stimme voll und ganz zu | Weiß nicht/trifft nicht zu) (i.e.: Zustimmung 1-5 und offen)
  - 3.5.1. …sich die Corona-Situation in Deutschland insgesamt verschlechtert.
  - 3.5.2. …es in meiner Nähe einen Ausbruchsherd gibt.
  - 3.5.3. …es in meinem sozialen Umfeld (z.B. Freunde, Bekannte, Familie) einen Fall gibt
  - 3.5.4. …Corona in den Medien präsenter ist.
  - 3.5.5. Sonstiges: (Freitext)
- 3.6. Wenn Sie die Corona-Warn-App auf Ihrem Mobiltelefon öffnen, warum tun Sie das? Ich öffne die Corona-Warn-App... (Antwortoptionen: Zustimmungsgrad, wie unter 3.5)
  - 3.6.1. …aus allgemeinem Interesse an der App.
  - 3.6.2. …um meinen Corona-Status zu überprüfen.
  - 3.6.3. …um Informationen über Corona zu erhalten.
  - 3.6.4. …um einen positiven oder negativen Test einzugeben, falls ich getestet wurde.
  - 3.6.5. …um das Funktionieren der Corona-Warnung sicherzustellen.
  - 3.6.6. Sonstiges: (Freitext)
4. Nur falls 1.2 mit „ja" beantwortet wurde: Nutzungsgründe und -effekte der Corona-Warn-App
   - 4.1. Wie hat sich Ihre Einstellung zu Ihrem Risiko durch Corona seit Verwendung der Corona-Warn-App geändert? Mit der Corona-Warn-App... (Antwortoptionen: Zustimmungsgrad, wie unter 3.5)
     - 4.1.1. …habe ich insgesamt ein besseres Gefühl von Sicherheit.
     - 4.1.2. …fühle ich mich sicherer im öffentlichen Raum (z.B. im Bus oder Zug, in Geschäften oder Restaurants).
     - 4.1.3. …fühle ich mich auf Situationen mit möglicherweise höherem Corona-Risiko besser vorbereitet.
     - 4.1.4. …denke ich, dass ich eher gesund bleiben kann.
     - 4.1.5. Sonstiges: (Freitext)
   - 4.2. Welche Beweggründe haben Sie für die Verwendung der Corona-Warn-App? Ich nutze die Corona-Warn-App, ... ... (Antwortoptionen: Zustimmungsgrad, wie unter 3.5)
     - 4.2.1. …um mich selbst vor Corona zu schützen.
     - 4.2.2. …um meine Familie vor Corona zu schützen.
     - 4.2.3. …um beizutragen, die Pandemie zu beenden.
     - 4.2.4. …weil Freunde oder Verwandte mich überzeugt haben.
     - 4.2.5. …weil öffentliche Werbung oder Nachrichten mich überzeugt haben.
     - 4.2.6. …weil Personen aus meinem Umfeld mich dazu gedrängt haben.
     - 4.2.7. Sonstiges: (Freitext)
   - 4.3. Was ist Ihnen zum Thema Datenschutz der Corona-Warn-App bekannt? (Antwortoptionen: Zustimmungsgrad, wie unter 3.5)
     - 4.3.1. Ich verstehe, welche Daten erhoben werden.
     - 4.3.2. Mir ist klar, welche Daten ausgetauscht werden.
     - 4.3.3. Ich habe Vertrauen, dass die App meine Daten angemessen schützt.
     - 4.3.4. Ich verstehe, auf welche Weise mein Infektionsrisiko gemessen wird.
     - 4.3.5. Ich habe ausreichend Kontrolle darüber, was mit meinen Daten geschieht.

5. Demografische Angaben (für alle Teilnehmer)
    5.1. Wie alt sind Sie? (Angabe in Jahren)
    5.2. Welchem Geschlecht ordnen Sie sich zu? (männlich, weiblich, divers)
    5.3. In welchem Land haben Sie Ihren Hauptwohnsitz? Antwortoptionen: Deutschland, Dänemark, Niederlande, Belgien, Luxemburg, Frankreich, Schweiz, Österreich, Tschechien, Polen Anderes Land.
    5.4. Nur falls bei 5.3 „Deutschland" ausgewählt wurde: Bitte geben Sie die ersten 3 Stellen der Postleitzahl Ihres Wohnortes an.
    5.5. Was ist ihre höchste, abgeschlossene Ausbildung? Antwortoptionen: Ohne Abschluss, Hauptschulabschluss, Mittlere Reife, Abitur oder Gleichwertiges, Studium, Doktor-Grad, Sonstiges.
    5.6. Benutzen Sie andere Tracking-Apps und Geräte? Antwortoptionen: Ja nutze ich, Nutze nicht mehr, Noch nie genutzt.
        5.6.1. Corona-Datenspende-App („RKI Datenspende").
        5.6.2. Eine Smartwatch wie z.B. Apple Watch oder Galaxy Watch
        5.6.3. Aktivitätstracker für tägliche Schritte und Aktivität.
        5.6.4. Sportuhren oder Apps zum Monitoring von sportlichen Aktivitäten wie Laufen, Schwimmen, Fahrradfahren u.a.
        5.6.5. Eine mit dem Mobiltelefon oder dem Internet verbundene Waage.
        5.6.6. Ein mit dem Mobiltelefon oder dem Internet verbundenes Blutzuckermessgerät.
        5.6.7. Ein mit dem Mobiltelefon oder dem Internet verbundenes Blutdruckmessgerät.
        5.6.8. Weitere Tracker und Apps (Freitext)
    5.7. Wie ist - unabhängig von der Corona Warn-App - Ihre persönliche Einstellung zu Datenschutz im Allgemeinen, zur Bedrohung für Sie durch Corona, und zu neuen Technologien im Allgemeinen? (Antwortoptionen: Zustimmungsgrad wie unter 3.5)
        5.7.1. Für mich ist es vor allem wichtig, meine persönlichen Daten vor dem Zugriff von Internetunternehmen zu schützen.
        5.7.2. Im Vergleich zu anderen Themen, die mir ein Anliegen sind, ist der Datenschutz sehr wichtig.
        5.7.3. Ich sorge mich oft darüber, mit dem COVID-19-Virus infiziert zu werden.
        5.7.4. Wenn ich vom Coronavirus höre, denke ich oft, dass ich mich mit ihm infiziert haben könnte.
        5.7.5. Ich fühle mich sicher beim Erlernen neuer Technologien.
        5.7.6. Ich habe ein gutes Verständnis für Technologie und Computer.

# 8 Danksagung

# 9 Literaturverzeichnis

Agarwal, James, Naresh K. Malhotra, und Sung S. Kim. 2004. „Internet Users' Information Privacy Concerns (IUIPC): The Construct, the Scale, and a Causal Model." *Information Systems Research*, Dezember: 336-355.

Heinssen, Robert K., Carol R. Glass, und Luanne A. Knight. 1987. „Assessing computer anxiety: Development and validation of the Computer Anxiety Rating Scale." *Computers in Human Behavior*, 49-59.

Salkovskis, P., K. Rimes, H. Warwick, und D. Clark. 2002. „The Health Anxiety Inventory: Development and validation of scales for the measurement of health anxiety and hypochondriasis." *Psychological Medicine*, 843-853.

Trang, Simon, Manuel Trenz, Welf H. Weiger, Monideepa Tarafdar, und Christy M.K. Cheung. 2020. „One app to trace them all? Examining app specifications for mass acceptance of contact-tracing apps." *European Journal of Information Systems.*